\documentclass[final,leqno,onefignum,onetabnum]{siamltex1213}

\title{Simulating Flexible Fiber Suspensions using a Scalable Immersed
  Boundary Algorithm \footnotemark[1]}  
\author{Jeffrey K. Wiens\footnotemark[2] \and
  John M. Stockie\footnotemark[2]} 

\usepackage{vmargin}

\usepackage{color}
\usepackage{colortbl,xcolor}
\usepackage{graphicx}
\usepackage{subfigure}
\usepackage{cancel}
\usepackage{amsfonts}
\usepackage{amsmath,amssymb}
\usepackage{pifont}
\usepackage{booktabs}
\usepackage{rotating}
\usepackage{algorithm}
\usepackage{algorithmicx}
\usepackage{algpseudocode}
\usepackage{multirow}
\usepackage{bbm}
\usepackage{calc}

\newcommand{\bsol}{\begin{proof}[Solution]}
\newcommand{\esol}{\end{proof}}
\newcommand{\bq}{\begin{equation}}
\newcommand{\eq}{\end{equation}}

\newcommand{\dt}{\Delta t}
\newcommand{\Laplacian}{\nabla^2}

\newcommand{\bs}[1]{\boldsymbol{#1}}
\newcommand{\half}{\frac{1}{2}}
\newcommand{\brac}[1]{\left(#1 \right)}

\newcommand{\pd}[2]{\frac{\partial #1}{\partial #2}}

\newcommand{\leavethisout}[1]{}

\newcommand{\GIBds}{\Delta s}
\newcommand{\ds}{\Delta s}
\newcommand{\EE}{\text{e}}

\newcommand{\veldev}{{\mathcal E}_{\text{rel}}}

\newcommand{\mystar}{\ast}

\newcommand{\Reynolds}{\mbox{\itshape Re}}

\newcommand{\orbiti}{{\sc I}}
\newcommand{\orbitii}{{\sc II}}
\newcommand{\orbitiii}{{\sc III}}
\newcommand{\orbitiiia}{{\sc IIIa}}
\newcommand{\orbitiiib}{{\sc IIIb}}
\newcommand{\orbitiv}{{\sc IV}}

\begin{document}
\maketitle

\renewcommand{\thefootnote}{\fnsymbol{footnote}}

\footnotetext[1]{We acknowledge support from the Natural Sciences and
  Engineering Research Council of Canada (NSERC) through a Postgraduate
  Scholarship (JKW) and a Discovery Grant (JMS). The numerical
  simulations in this paper were performed using computing resources
  provided by WestGrid and Compute Canada.}

\footnotetext[2]{Department of Mathematics, Simon Fraser University,
  Burnaby, BC, Canada, V5A 1S6 (\email{jwiens@sfu.ca},
  \email{jstockie@sfu.ca}).}

\renewcommand{\thefootnote}{\arabic{footnote}}

\slugger{sisc}{xxxx}{xx}{x}{x--x}

\begin{abstract}
  We present an approach for numerically simulating the dynamics of
  flexible fibers in a three-dimensional shear flow using a scalable
  immersed boundary (IB) algorithm based on Guermond and Minev's
  pseudo-compressible fluid solver.  The fibers are treated as
  one-dimensional Kirchhoff rods that resist stretching, bending, and
  twisting, within the {\it generalized IB} framework.  We perform a
  careful numerical comparison against experiments on single fibers
  performed by S.~G.~Mason and co-workers, who categorized the fiber
  dynamics into several distinct orbit classes.  We show that the orbit
  class may be determined using a single dimensionless parameter for low
  Reynolds flows.  Lastly, we simulate dilute suspensions containing up
  to hundreds of fibers using a distributed-memory computer cluster.
  These simulations serve as a stepping stone for studying more complex
  suspension dynamics including non-dilute suspensions and aggregation
  of fibers (also known as flocculation).
\end{abstract}



\pagestyle{myheadings}
\thispagestyle{plain}
\markboth{J.~K. WIENS AND J.~M. STOCKIE}{%
  IB SIMULATIONS OF FLEXIBLE FIBER SUSPENSIONS%
  }

\section{Introduction}

The behaviour of long, flexible fibers in a suspension plays an
important role in many applications, including pulp and paper
manufacture, polymer melts, and fiber-reinforced composite
materials~\cite{keshtkar-etal-2009, switzer-klingenberg-2003}.  The
dynamics of such suspensions depend heavily on the shape and flexibility
of the individual fibers as well as the interactions between fibers.
Because of the complexity of the fiber motion in suspensions, many
researchers have developed numerical methods that afford valuable
insight into both individual fiber dynamics and the resulting aggregate
suspension rheology~\cite{Joung2001, Petrie1999,
  switzer-klingenberg-2003}.  These simulations can complement physical
experiments by providing information that is not easily obtained through
direct measurement.

In this paper, we develop an approach for simulating a suspension of
flexible fibers that is based on the immersed boundary (IB)
method~\cite{Peskin2002}, which is a mathematical framework originally
developed by Peskin~\cite{Peskin1972} to capture the two-way interaction
between a fluid and an immersed deformable structure.  Here, the fluid
deforms the elastic structure while the structure exerts forces onto the
fluid.  The IB method has been used to study a wide variety of
biological and engineering applications including blood flow through
heart valves~\cite{Griffith2009,Peskin1972}, cell growth and
deformation~\cite{Rejniak2007}, jellyfish locomotion~\cite{Hamlet2011},
evolution of dry foams~\cite{Kim2012} and parachute
aerodynamics~\cite{Kim2009}.

We treat the flexible fibers as one-dimensional Kirchhoff
rods~\cite{Dill1992} described using the {\it generalized IB framework}
developed by Lim et al.~\cite{Lim2008}.  In this approach, the fibers
are represented as 1D space curves using a moving Lagrangian coordinate,
wherein at each Lagrangian point an orthonormal triad of vectors
describes the orientation and ``twist state'' of the rod.  This permits
the fiber to generate not only a force but also a torque that is applied
to the surrounding fluid.

The primary objective of this paper is to develop an efficient
methodology for simulating suspensions containing a large number of
flexible fibers. Since solving the full fluid-structure interaction
problem comes at the expense of additional computational work, the
underlying parallel algorithm is purposely designed to scale efficiently
on distributed-memory computer clusters.  This permits non-dilute
suspensions to be simulated efficiently by spreading the work over
multiple processors. The numerical algorithm is based on the work of
Wiens and Stockie~\cite{Wiens2013} who implemented a pseudo-compressible
fluid solver developed by Guermond and
Minev~\cite{Guermond2010,Guermond2011} in the IB framework.  We extend
this original algorithm to use the Eulerian--Lagrangian discretization
employed by Griffith and Lim~\cite{Griffith2012-2} which employs a
predictor-corrector procedure to evolve the immersed boundary.  
Here, two separate force spreading and velocity interpolation steps
are applied at each time step which improves the spatial convergence rate of the method.

We begin in Section~\ref{sec:Background} by reviewing theoretical and
experimental results in the literature pertaining to the hydrodynamics
of suspensions containing flexible fibers, as well as discussing several
prominent computational approaches.  In Sections~\ref{sec:Equations}
and~\ref{sec:NumericalMethod}, we state the governing equations
underlying our IB model for fluid-fiber interaction, as well as the
numerical algorithm used to approximate these equations. In
Section~\ref{sec:Results}, we present simulations of fiber dynamics in
both single- and multi-fiber systems, and compare these results to
previously published experimental work.

\section{Background: Pulp Fibers}
\label{sec:Background}

\subsection{Theory and Experiments}

Theoretical investigations of the dynamics of fibers in a shear flow
date back to Jeffery in the 1920s~\cite{Jeffery1922}, who derived an
analytical solution for the motion of a single rigid, neutrally-buoyant
ellipsoidal particle immersed in an incompressible Newtonian fluid
(specifically, in a Stokes flow).  Jeffery found that such a fiber
rotates with a well-defined periodic orbit having constant period but
non-uniform angular velocity.  It was later shown by
Bretherton~\cite{Bretherton1962} that Jeffery's analytical solution
could be extended to more general axisymmetric particles with
non-elliptical cross-sections by replacing the ellipsoidal aspect ratio
$a_r$ by an effective aspect ratio $a_r^\mystar$.

Although the theory for rigid fiber dynamics is relatively
well-developed, far less is known about fibers that experience
significant bending. For this reason, experimental observations are of
critical importance in understanding the dynamics and rheology of
suspensions containing flexible fibers. Unlike rigid fibers, flexible
fibers undergo a much wider and richer range of motion when subjected to
a background linear shear flow given with velocity field
$\bs{u}=(Gy,0,0)$. This problem was studied in the pioneering work of
Mason and co-workers~\cite{Arlov1958,Forgacs1959,Forgacs1958} who
categorized the fiber dynamics into several distinct orbit classes.
When motions are confined to the $xy$-plane, fiber dynamics fall into
one of four orbit classes -- rigid, springy, flexible, and complex rotations --
which are illustrated in Table~\ref{Table:FiberRotations}.  The
experiments of Mason et al.~involved primarily synthetic fibers (made of
rayon and dacron) immersed in highly viscous fluids (such as corn syrup)
although their original motivation was the application to natural wood
pulp fiber suspensions.

\begin{table}[ht!bp]\centering\small
  \caption{Two-dimensional orbit classes for flexible fibers whose
    unstressed state is intrinsically straight.  Adapted from Forgacs et
    al.~\cite{Forgacs1958}.}  
  \begin{tabular}{ll c}\toprule
    & Orbit Class & \multicolumn{1}{c}{} \\
    \midrule
       \orbiti    & Rigid rotation & \includegraphics[height=.6in]{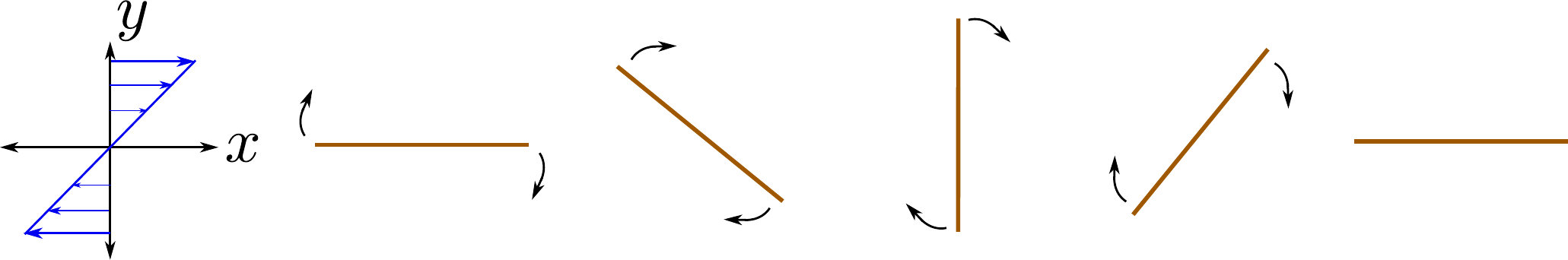} \\
           \midrule
       \orbitii   & Springy rotation & \includegraphics[height=.6in]{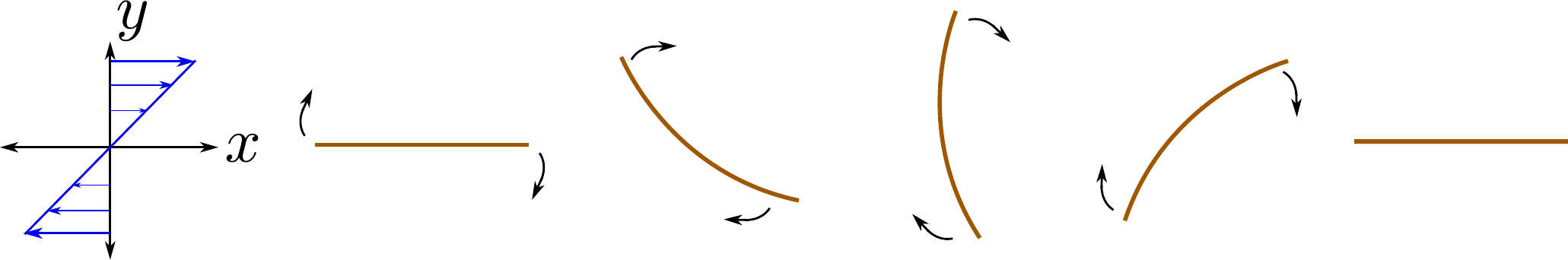} \\
           \midrule
       \orbitiiia & Loop or S~turn & \includegraphics[height=.6in]{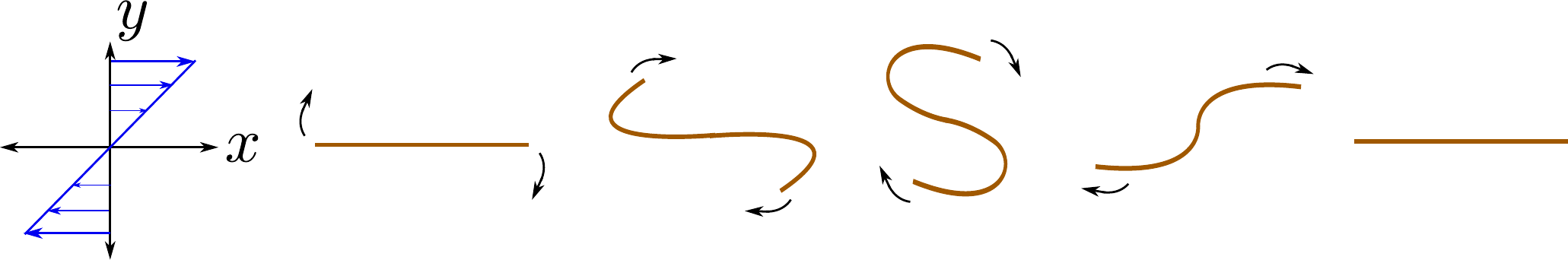} \\
       \cmidrule(r){3-3}
       \orbitiiib & Snake turn & \includegraphics[height=.6in]{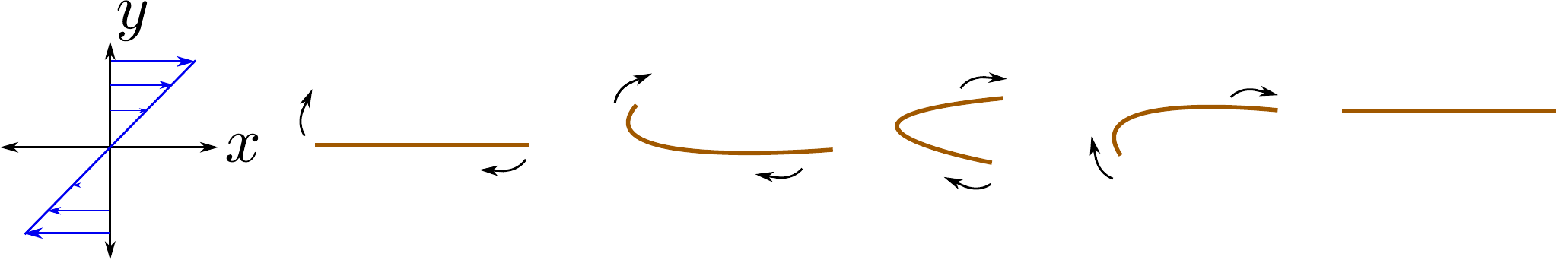} \\
           \midrule
       \orbitiv   & Complex rotation & \includegraphics[height=.6in]{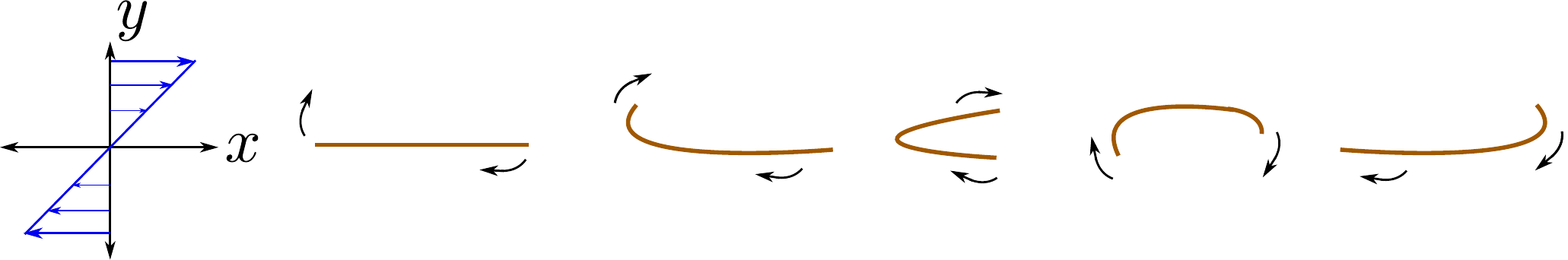} \\
    \bottomrule
  \end{tabular}
  \label{Table:FiberRotations}
\end{table}

These experiments on fiber suspensions demonstrate that varying either
the hydrodynamic drag force or the fiber flexibility governs the
transition between the various planar orbit classes.  In
class~\orbiti~orbits, the fiber remains rigid and rotates as predicted
by Jeffery's equation.  When a small flexibility is introduced into the
fiber, it undergoes a springy rotation (class~\orbitii) in which it
bends into a shallow arc as it rotates outside the horizontal plane of
shear. When the fiber flexibility is increased, it experiences
significant deformations that take the form of {\it S~turns}
(class~\orbitiiia) or {\it snake turns} (class~\orbitiiib).  Note that
S~turns require a high degree of initial symmetry so that snake turns
are actually far more prevalent in actual
suspensions~\cite{Arlov1958,Forgacs1959}.  When the fiber flexibility is
increased even further, the fiber may never straighten out as it returns
to the horizontal, in which case the orbit is classified as a complex
rotation (class~\orbitiv).  For the largest values of flexibility
encountered in thread-like synthetic fibers, the fiber can transition
beyond the class of complex rotations and undergo convoluted
self-intersections as observed by Forgacs and Mason~\cite{Forgacs1959}
in experiments.

In many cases, the fiber rotation is not constrained to the $xy$-plane
but instead undergoes a genuinely three-dimensional orbit that protrudes
or ``buckles'' out along the $z$-direction, although the $xy$-projection
of the fiber may still belong to one of the planar orbit classes
\orbiti--\orbitiv\ described above.  Note that real suspensions such as
wood pulp also contain irregularly-shaped fibers that are either
intrinsically curved or contain kinks or other non-uniformities;
consequently, fiber orbital dynamics in such suspensions are not
necessarily confined to these idealized orbit classes.  Indeed, the
experiments of Arlov et al.~\cite{Arlov1958} were used to classify a
much broader class of genuinely three-dimensional orbits for wood pulp
fibers having an intrinsic curvature.

We close this discussion by defining a dimensionless parameter that can
be used to conveniently classify and predict the orbit class to which a
specific fiber belongs.  For low Reynolds number flows (with $\Reynolds
\lessapprox 1$), the hydrodynamic drag force experienced by a fiber is
proportional to
\begin{gather}
  \label{eq:DragForce}
  F_d = \mu G D,
\end{gather}
where $\mu$ is the fluid viscosity, $G$ is the shear rate, and $D$ is
the diameter of the fiber~\cite{White2006}.  By balancing this drag
force with the corresponding fiber bending force, a
single dimensionless parameter can be derived that captures the fiber
flexibility~\cite{Stockie1997}
\begin{gather}
  \label{eq:DimensionlessShearRate}
  \chi = \frac{\mu D G L^3}{EI},
\end{gather}
where $L$ is the fiber length, $E$ is Young's modulus of the material,
and $I$ is moment of area in the plane of bending.  The parameter $\chi$
may also be interpreted as a ratio of fiber deflection to fiber length.
In a series of 2D numerical simulations~\cite{Stockie1998}, the
parameter $\chi$ was shown to provide a useful measure of fiber
flexibility that characterizes each orbit class over a wide range of
fluid and fiber parameters.  This dimensionless flexibility parameter
has also appeared in the computational studies of Ross and
Klingenberg~\cite{Ross1997} (where they referred to it as a
dimensionless shear rate) and Wherrett et~al.~\cite{Wherrett1997} (where
$\chi^{-1}$ is called a bending number).

\subsection{Overview of Computational Approaches}

A popular class of numerical methods for simulating flexible fibers is
the so-called {\it bead models} in which a flexible fiber is treated as
a string of rigid beads that are linked together by flexible
connectors. This approach originated with the work of Yamamoto and
Matsuoka~\cite{Yamamoto1993} who treated fibers as chains of bonded
spheres that are free to stretch, bend and twist relative to each other.
Their approach was extended by Ross and Klingenberg~\cite{Ross1997} who
modelled fibers as chains of rigid prolate spheroids connected by ball
and socket joints.  The dynamics of the bead network are governed by
Newton's laws through a balance of linear and angular momentum that
incorporates the hydrodynamic and interparticle forces acting on each
bead.  More recently, Klingenberg's group has validated their model
results against experiments for single fiber dynamics~\cite{Skjetne1997}
as well as developing a multi-fiber extension that has been used to
simulate flocculation~\cite{Switzer2004}.

A significant shortcoming of Klingenberg's model and related
variants~\cite{Wang2006,Wherrett1997,Yamamoto1993} is that they fail to
capture the full fluid-structure interaction in fiber suspensions.
Although their approach does include the hydrodynamic force exerted by
the fluid on the fiber, the fiber does not itself exert any force back
onto the fluid; therefore, the fluid is a passive medium that obviously
neglects any of the complex fluid dynamics that must occur in the region
immediately adjacent to a dynamically deforming fiber.  Several recent
bead-type models have attempted to address this limitation, for example
Wu and Aidun who proposed a model for rigid~\cite{Wu2010} and
flexible~\cite{Wu2010-2} fibers that incorporates the full
fluid-structure interaction using a Lattice Boltzmann approach.
Similarly, Lindstr{\"o}m and Uesaka proposed an alternative model for
rigid~\cite{Lindstrom2009} and
flexible~\cite{Lindstrom2007,Lindstrom2008} fibers that uses the
incompressible Navier--Stokes equations to model the fluid.

A completely different approach for capturing flexible fiber dynamics is
based on the slender body theory~\cite{Batchelor1970} which exploits
approximations to the governing equations based on a small fiber aspect
ratio. This is the approach taken by Tornberg and
Shelley~\cite{TornbergShelley2004} who studied flexible filaments in a
Stokes flow by deriving a system of one-dimensional integral equations.
They solved these integral equations numerically using a second-order
method that also captures interactions between multiple fibers. This
approach has been further extended by Li et al.~\cite{Li2013} who used a
similar methodology to investigate the problem of sedimentation (or
settling) of flexible fibers.  Unlike the bead models described earlier,
this slender-body approach cleanly separates the fiber model from its
numerical treatment, which makes the model more amenable to mathematical
analysis and also permits the numerical discretization to be
independently tested through convergence studies.  Furthermore, because
the fluid has been simplified by assuming a Stokes flow regime, these
slender-body discretizations do not require a fluid grid because of the
availability of numerical methods based on Green's-function solutions
that greatly reduce the computational complexity. The only significant
disadvantage of this approach, beside the Stokes flow restriction, is
that there are as yet no results that incorporate any effects of fiber
twist~\cite{Olson2013}.

An alternative approach that permits simulating flexible fibers immersed
in higher Reynolds flows is the immersed boundary method. This is the
approach taken by Stockie and Green~\cite{Stockie1998} who simulated a
single flexible fiber in two dimensions using a simple representation of
the fiber in terms of spring-like forces that resist stretching and
bending.  Stockie~\cite{Stockie2002} later extended these results to 
a single 3D wood pulp fiber using a much more detailed and realistic
model that explicitly captures the interwoven multi-layer network of
cellulose fibrils making up the wood cell wall. More recently, Nguyen
and Fauci studied diatom chains using the IB method with a similarly
detailed fiber model~\cite{Nguyen2014}.  The IB method properly captures
the full interaction between the fluid and immersed structure by
including the appropriate no-slip boundary condition along the fiber,
although it does come at an additional cost.  First of all, in
comparison with slender-body models, the fluid solver portion of the IB
algorithm can be significantly more expensive because it solves the
Navier-Stokes equations on a finite difference grid. Secondly, because
the IB method aims to capture the detailed fluid flow around the fiber,
the fluid grid needs to be adequately refined in order to resolve
details on the order of the fiber diameter, which in turn places
practical limitations on the fiber aspect ratio that can be computed.
Thirdly, a detailed characterization of the structure of a
three-dimensional fiber such as in~\cite{Nguyen2014,Stockie2002}
typically requires thousands of IB points to resolve and is therefore
computationally impractical for simulating semi-dilute suspensions of
multiple fibers.

In this paper, we apply the IB approach to simulate flexible fibers, and
we have chosen to treat each fiber instead as a one-dimensional
Kirchhoff rod that resists stretching, bending and twisting, as
described in the generalized IB method of Lim et al.~\cite{Lim2008}.
Additionally, we employ a highly scalable implementation of the
generalized IB algorithm~\cite{Wiens2013} that spreads the computational
work over a large number of processors, thereby permitting us to
simulate hydrodynamic interactions in suspensions containing large
numbers of flexible fibers.

\section{Governing Equations}
\label{sec:Equations}

Consider a Newtonian, incompressible fluid that fills a rectangular
domain $\Omega$ having dimensions $H_x \times H_y \times H_z$ and whose
state is specified using Eulerian coordinates $\bs{x}=(x,y,z)$.
Immersed within the fluid is a neutrally-buoyant elastic fiber of length
$L$.  The fiber is described by a one-dimensional space curve $\Gamma
\subset \Omega$, parameterized by the Lagrangian coordinate $s\in
[0,L]$.  The spatial configuration of the rod at time $t$ is given in
parametric form as $\bs{x}=\bs{X}(s,t)$ and its orientation and ``twist
state'' are defined in terms of the orthonormal triad of vectors
$\{\bs{D}^1(s,t),\bs{D}^2(s,t),\bs{D}^3(s,t)\}$, where the third triad
vector $\bs{D}^3$ remains tangent to the space curve $\bs{X}$.  Note
that because of numerical considerations (described shortly),
$\bs{D}^3(s,t)$ is not exactly tangent to the space curve $\bs{X}$ but
is rather penalized in a way that it is only approximately in the
tangential direction.

The fluid velocity $\bs{u}(\bs{x},t)$ and pressure $p(\bs{x},t)$ at
location $\bs{x}$ and time $t$ are governed by the incompressible
Navier--Stokes equations
\begin{gather}
  \label{eq:NSE}
  \rho \brac{\pd{\bs{u}}{t} + \bs{u}\cdot\nabla\bs{u}} + \nabla p = \mu 
  \Laplacian \bs{u} + \bs{f} + \half \nabla \times \bs{n}, 
  \\
  \label{eq:incompressible}
  \nabla \cdot \bs{u} = 0,
\end{gather}
where $\rho$ is the fluid density and $\mu$ is the dynamic viscosity
(both constants).  The {\it Eulerian} force and torque densities,
$\bs{f}$ and $\bs{n}$, are written as
\begin{align}
  \label{eq:force}
  \bs{f}(\bs{x},t) =& \int\limits_\Gamma \bs{F}(s,t) \, \Phi_w(\bs{x} -
  \bs{X}(s,t)) \,ds \qquad  \text{and} \\
  \label{eq:torque}
  \bs{n}(\bs{x},t) =& \int\limits_\Gamma \bs{N}(s,t) \, \Phi_w(\bs{x} -
  \bs{X}(s,t)) \,ds,
\end{align}
wherein the integrals spread the {\it Lagrangian} force and torque
densities, $\bs{F}(s,t)$ and $\bs{N}(s,t)$, onto points in the fluid.
The interaction between Eulerian and Lagrangian quantities is mediated
using the smooth kernel function 
\begin{gather}
  \label{eq:fulldiscretekernel}
  \Phi_w(\bs{x}) = \frac{1}{w^3} \, \phi \brac{ \frac{x_1}{w} } \phi
  \brac{ \frac{x_2}{w} } \phi
  \brac{ \frac{x_3}{w} },
\end{gather}
where
\begin{gather}
  \label{eq:discretekernel}
  \phi(r) = 
    \begin{cases}
      \frac{1}{8}(3-2|r| + \sqrt{1+4|r|-4r^2}) & 
      \text{if $0 \leq |r| < 1$}, \\
      \frac{1}{8}(5-2|r| - \sqrt{-7+12|r|-4r^2}) & 
      \text{if $1 \leq |r| < 2$}, \\
      0 & \text{if $2 \leq |r|$}.
    \end{cases}
\end{gather}
Here, $w$ represents an effective thickness of the rod which is set to
some multiple of the fluid mesh width $h$; that is, $w=C h$ for some
integer multiple $C \in \mathbb{Z}^+$.  Note that if $w=h$, the kernel
$\Phi_w(\bs{x})$ is identical to the discrete delta function employed in
many immersed boundary methods~\cite{Griffith2005,Lai2000,Mori2008}.

The rod is modeled as a Kirchhoff rod~\cite{Dill1992} using the
generalized immersed boundary framework of Lim~\cite{Lim2008}.
Balancing linear and angular momentum yields the Lagrangian force and
torque densities
\begin{align}
  \label{eq:linearmomentumbalance}
  \bs{F} =&~ \pd{\bs{F}^{\text{rod}}}{s} \text{, } \\
  \label{eq:angularmomentumbalance}
  \bs{N} =&~ \pd{\bs{N}^{\text{rod}}}{s} + \pd{\bs{X}}{s} \times
  \bs{F}^{\text{rod}} \text{, } 
\end{align}
in terms of the internal force $\bs{F}^{\text{rod}}(s,t)$ and moment
$\bs{N}^{\text{rod}}(s,t)$ transmitted across a segment of the rod.
Internal quantities are expanded in the basis $\{ \bs{D}^1, \bs{D}^2,
\bs{D}^3\}$ as
\begin{align}
  \label{eq:rodforce}
  \bs{F}^{\text{rod}} = &~  F^1 \bs{D}^1 + F^2 \bs{D}^2 + F^3 \bs{D}^3,\\
  \label{eq:momentforce}
  \bs{N}^{\text{rod}} = &~  N^1 \bs{D}^1 + N^2 \bs{D}^2 + N^3 \bs{D}^3,
\end{align}
where the coefficient functions are defined by the constitutive
relations 
\begin{align}
  \label{eq:momentconstitutive}
  N^1 =&~ a_1 \brac{\pd{\bs{D}^2}{s} \cdot \bs{D}^3 - \kappa_1}, & 
  N^2 =&~ a_2 \brac{\pd{\bs{D}^3}{s} \cdot \bs{D}^1 - \kappa_2}, &  
  N^3 =&~ a_3 \brac{\pd{\bs{D}^1}{s} \cdot \bs{D}^2 - \tau},\\
  \label{eq:forceconstitutive}
  F^1 =&~ b_1   \brac{\bs{D}^1 \cdot \pd{\bs{X}}{s}}, & 
  F^2 =&~ b_2  \brac{\bs{D}^2 \cdot \pd{\bs{X}}{s}}, & 
  F^3 =&~ b_3 \brac{\bs{D}^3 \cdot \pd{\bs{X}}{s} - 1}.
\end{align}
Equations~\eqref{eq:momentconstitutive} incorporate the resistance of
the rod to bending and twisting, with $a_1$ and $a_2$ being the bending
moduli (about axes $\bs{D}^1$ and $\bs{D}^2$ respectively) while $a_3$
is the twisting modulus.  The constants
$\brac{\kappa_1,~\kappa_2,~\tau}$ define the intrinsic twist vector of
the rod where $\kappa := \sqrt{\kappa_1^2 + \kappa_2^2}$ is the
intrinsic curvature and $\tau$ is the intrinsic twist in the stress-free
configuration.  The remaining force terms~\eqref{eq:forceconstitutive}
act to keep the triad vector $\bs{D}^3$ approximately aligned with the
tangent curve $\partial \bs{X} / \partial s$ and also penalize any
stretching of the rod from its equilibrium configuration.  Accordingly,
the generalized IB method can be viewed as a type of penalty method in
which the rod is only approximately inextensible and approximately
aligned with the orthonormal triad, and the constants $b_1$, $b_2$ and
$b_3$ play the role of penalty parameters.

The final equations required to close the system are evolution equations
for the rod configuration and triad vectors
\begin{align}
  \label{eq:membrane}
  \pd{\bs{X}}{t}(s,t) =& ~\bs{U}(s,t)\text{, } \\
  \label{eq:twisting}
  \pd{\bs{D}^\alpha}{t}(s,t) =&~ \bs{W}(s,t) \times \bs{D}^\alpha(s,t) \text{,}
\end{align}
where $\alpha=1,2,3$, and $\bs{U}(s,t)$ and $\bs{W}(s,t)$ are the linear
and angular velocities along the axis of the rod respectively.  These
equations require that the rod translate and rotate according to the
local average linear and angular velocity of the fluid, and are
interpolated in the standard IB fashion as
\begin{align}
  \label{eq:LinVel}
  \bs{U}(s,t) =&~ \int\limits_\Omega
  \bs{u}(\bs{x},t) \,\Phi_w(\bs{x}-\bs{X}(s,t)) \, d\bs{x} \text{, } \\
  \label{eq:AngVel}
  \bs{W}(s,t) =&~ \half \int\limits_\Omega
  \nabla \times \bs{u}(\bs{x},t) \,\Phi_w(\bs{x}-\bs{X}(s,t)) \, d\bs{x}.
\end{align}
By using the same kernel function $\Phi_w$ as in
\eqref{eq:force}--\eqref{eq:torque}, we ensure that energy is conserved
during the Eulerian--Lagrangian interactions~\cite{Lim2008}.

\subsection{Problem Geometry and Initial Conditions}
\label{sec:InitialConditions}

\begin{figure}[!htbp]
  \begin{center}
    \includegraphics[width=0.8\textwidth]{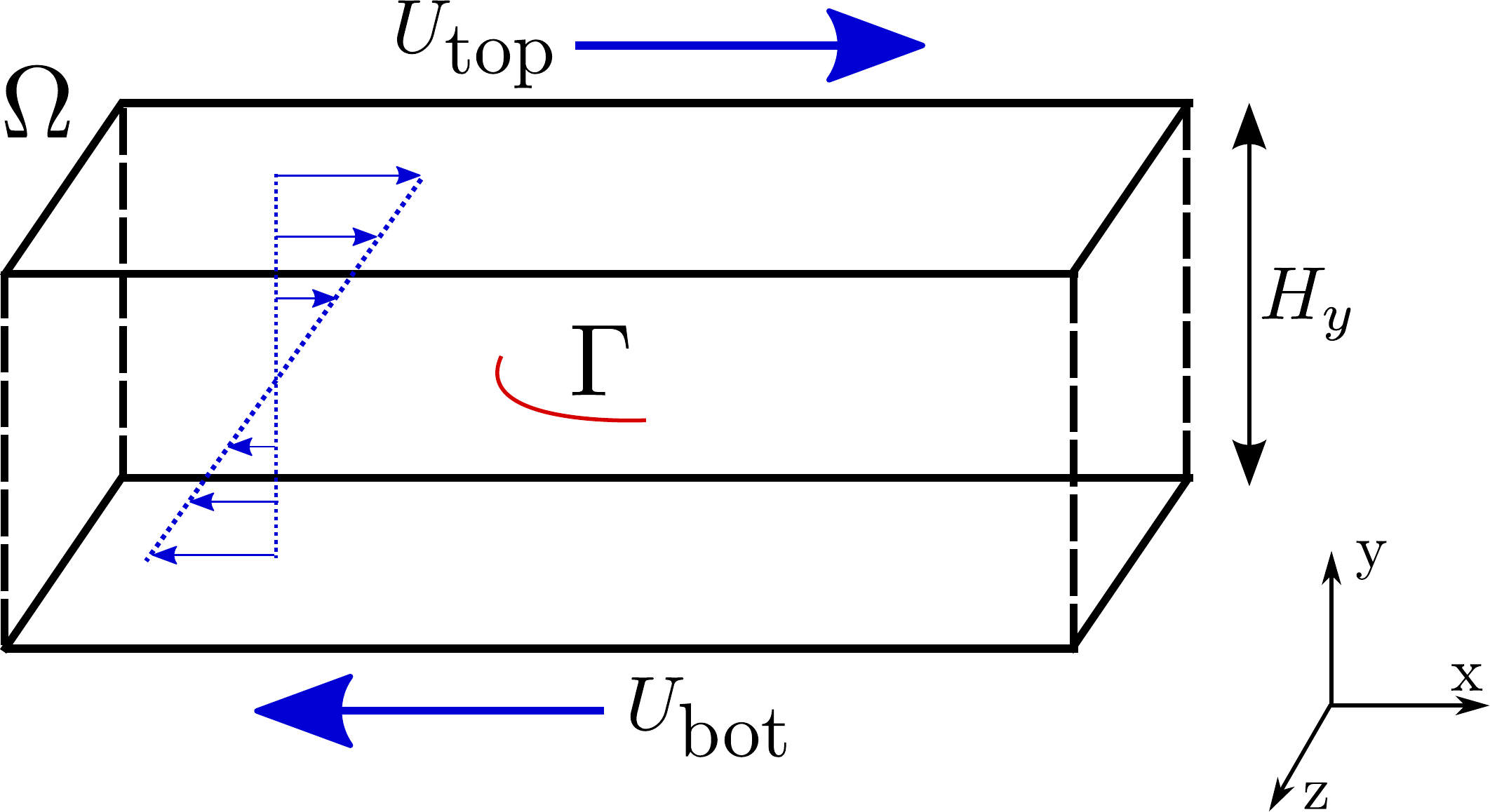}
    \caption{Problem geometry for a single fiber $\Gamma$ located at the
      center of a periodic, rectangular channel $\Omega$ of dimension
      $H_x\times H_y\times H_z$.  A planar shear flow is generated by
      forcing the top and bottom walls to move with constant velocities
      $\pm U_{top}$.}
  \label{fig:ShearFluidDomain}
  \end{center}
\end{figure}

The problem geometry is illustrated in
Figure~\ref{fig:ShearFluidDomain}, showing a fiber $\Gamma$ immersed in
a rectangular fluid domain $\Omega$. Periodic boundary conditions are
imposed on the fluid in the $x$- and $z$-directions, while the fluid is
sheared in the vertical ($y$) direction.  The shear flow is induced by
imparting a horizontal motion to the top and bottom boundaries, with the
top wall moving at speed $U_{\text{top}}$ and the bottom wall in the
opposite direction at speed $U_{\text{bot}}$.  In practice, we impose
$U_{\text{top}}=U_{\text{bot}} := U$ and set the initial fluid velocity
to the linear shear profile $\bs{u}(\bs{x},0) = \brac{G(y-H_y/2), 0, 0}$
that would develop in the absence of the fiber, with shear rate $G =
2U/H_y$.
The fiber of length $L$ is placed at the center of the fluid domain
which is specified by the constant $\bs{X}_0$, and we consider three
different initial configurations for the fiber:
\begin{description}
\item[\normalfont\itshape Configuration 1.] The fiber is initially
  straight and is parameterized by
  \begin{align*}
    \bs{X}(s,0) &= \brac{ (\epsilon_0+1) s,~0,~ 0} + \bs{X}_0,
    \\
    \bs{D}^1(s,0) &= \brac{0,~1,~0},
    \\
    \bs{D}^2(s,0) &= \brac{0,~0,~1},
    \\
    \bs{D}^3(s,0) &= \brac{1,~0,~0},
  \end{align*}
  where $0 \leq s < L$ and $\epsilon_0$ is a perturbation parameter that
  initially stretches the fiber.
  
\item[\normalfont\itshape Configuration 2.] The fiber is curved in the $xy$-plane with
  \begin{align*}
    \bs{X}(s,0) &= \brac{ r_0 \cos(s/r_0 + \pi),~ r_0 \sin(s/r_0 + \pi),~ 0} + \bs{X}_0,
    \\
    \bs{D}^1(s,0) &= \brac{0,~0,~1},
    \\
    \bs{D}^2(s,0) &= \brac{\cos(s/r_0 + \pi),~\sin(s/r_0 + \pi),~0},
    \\
    \bs{D}^3(s,0) &= \brac{\sin(s/r_0),~\cos(s/r_0 + \pi),~0},
  \end{align*}
  where $\alpha_b r_0 \pi \leq s < \alpha_e r_0 \pi$, and $\alpha_b$ and
  $\alpha_e$ are constants with $0 \leq \alpha_b < \alpha_e \leq 1$.
  Here, the fiber is a segment of a circle of radius $r_0$ lying in the
  $xy$-plane and having length $L = (\alpha_e-\alpha_b) \pi r_0$.
  Choosing a sufficiently large radius $r_0$ generates fiber with
  small initial curvature.
  
\item[\normalfont\itshape Configuration 3.]  Similar to Configuration~2, 
  except that the fiber is curved in the $xz$-plane with
  \begin{align*}
    \bs{X}(s,0) &= \brac{ (\epsilon_0+r_0) \cos(s/r_0 ),~ 0,~
      (\epsilon_0+r_0) \sin(s/r_0 )} + \bs{X}_0, 
    \\
    \bs{D}^1(s,0) &= \brac{0,~-1,~0},
    \\
    \bs{D}^2(s,0) &= \brac{\cos(s/r_0 ),~0,~\sin(s/r_0)},
    \\
    \bs{D}^3(s,0) &= \brac{\sin(s/r_0+ \pi),~0,~\cos(s/r_0)},
  \end{align*}
  where $\alpha_b r_0 \pi \leq s < \alpha_e r_0 \pi$, and $\alpha_b$ and
  $\alpha_e$ are constants satisfying $0 \leq \alpha_b < \alpha_e \leq 1$.
\end{description}
For all three configurations, the rod has open ends so that boundary
conditions are required at $s = 0$ and $L$.  We assume that the internal
force and moment vanish at the endpoints, corresponding to
$\bs{F}^{\text{rod}}_{-1/2} = \bs{F}^{\text{rod}}_{N_s-1/2} = 0$ and
$\bs{N}^{\text{rod}}_{-1/2} = \bs{N}^{\text{rod}}_{N_s-1/2} = 0$, which
are consistent with the boundary conditions applied by
Lim~\cite{Lim2010}.

\section{Numerical Method}
\label{sec:NumericalMethod}

Here, we provide only a very brief overview of the numerical method used
to solve the governing equations, while a detailed description of the
method and its parallel implementation can be found 
in~\cite{Wiens2014,Wiens2013}.

When discretizing the governing equations we use two separate
computational grids, one each for the Eulerian and Lagrangian variables.
The fluid domain is divided into an $N_x \times N_y \times N_z$,
uniform, rectangular mesh where each cell has side length $h$. We employ
a \emph{marker-and-cell} (MAC) discretization~\cite{Harlow1965} wherein
the pressure is approximated at cell center points $\bs{x}_{i,j,k}$ for
$i,j,k = 0,1,\ldots,N-1$, while velocity components are located on cell
faces.  The Lagrangian variables are discretized at $N_s$
uniformly-spaced points denoted by $s_\ell = \ell \ds$ for $\ell=0, 1,
\ldots, N_s-1$ with $\ds = L/N_s$.  Since our current implementation is
restricted to periodic fluid domains, the top and bottom wall boundary
conditions are imposed by slightly increasing the size of the fluid
domain in the $y$-direction and introducing planes of {\it IB tether
  points} along $y=0$ and $H_y$ that are attached by very stiff springs to
points moving at the specified velocities $U_{\text{top}}$ and
$U_{\text{bot}}$.  We did this for convience only, since neither the
governing equations nor the fluid solver is restricted to periodic
domains.

The IB equations are approximated using a fractional-step method
described by Wiens and Stockie~\cite{Wiens2013} in which the calculation
of fluid variables is decoupled from that of the immersed boundary.  For
integrating the fluid equations, we use the pseudo-compressibility
method developed by Guermond and Minev~\cite{Guermond2010,Guermond2011},
which employs a directional-splitting strategy that reduces to a series
of one-dimensional tridiagonal systems.  These linear systems can be
solved efficiently on distributed-memory clusters by combining Thomas's
algorithm with a Schur-complement technique. 

When integrating the rod position and orthonormal triad vectors forward
in time, we use the predictor-corrector procedure devised by Griffith
and Lim~\cite{Griffith2012-2}.  This differentiates our numerical method
from the approach taken in~\cite{Wiens2013}, where an Adams--Bashforth
extrapolation was used to evolve the immersed boundary in time.
Although the predictor-corrector procedure introduces additional work,
this change is necessary in order to obtain second-order convergence
rates in space.

Lastly, the constitutive relations
\eqref{eq:linearmomentumbalance}--\eqref{eq:forceconstitutive} are
discretized in the same manner as in Lim et al.~\cite{Lim2008}, with the
main difference being in how the orthonormal triad vectors are
interpolated onto half Lagrangian steps $s_{\ell+\half} = (\ell+\half)
\ds$.  Here, we use the Rodrigues' rotation formula as described
in~\cite{Wiens2014} instead of taking the principal square root used by
Lim et al.~\cite{Lim2008}.

If we assume that the state variables are all known at time $t_n$, the
IB algorithm for a single time step $\dt$ proceeds as follows.
\begin{enumerate}
\setlength{\itemsep}{.5em}

\item Interpolate the linear and angular fluid velocities onto the
  rod using the the delta kernel $\Phi_w(\bs{x})$ to obtain $\bs{U}^n$
  and $\bs{W}^n$.

\item Predict the rod position $\bs{X}^{n+1,\mystar}$ and orthonormal
  triad vectors $(\bs{D}^\alpha)^{n+1,\mystar}$ at time
  $t_{n+1}=(n+1)\dt$ to first order for $\alpha=1,2,3$.

\item Calculate the {Lagrangian} force and torque densities,
  $\bs{F}$ and $\bs{N}$, at times $t_{n}$ and $t_{n+1}$ using the
  discretization employed by Lim et al.~\cite{Lim2008}.
              
\item Spread the Lagrangian force and torque densities just calculated
  onto fluid grid points.  Then approximate the {Eulerian} force
  and torque density, $\bs{f}^{n+\half}$ and $\bs{n}^{n+\half}$, at time
  $t_{n+\half}=(n+\half)\dt$ using an arithmetic average.

\item Integrate the incompressible Navier--Stokes equations to time
  $t_{n+1}$ using ($\bs{f}^{n+\half} + \half \nabla \times
  \bs{n}^{n+\half}$) as the external body force.

\item Correct the rod position $\bs{X}^{n+1}$ and orthonormal triad
  $(\bs{D}^\alpha)^{n+1}$ to second order.  This requires interpolating
  the linear and angular fluid velocity at time $t_{n+1}$ onto the rod
  location.
\end{enumerate}

\section{Numerical Results}
\label{sec:Results}

\subsection{Intrinsically Straight Fibers}
\label{sec:IntrinsicallyStraightFiber}

We begin by considering the behaviour of a single flexible fiber
immersed in a shear flow, where the equilibrium fiber state is
intrinsically straight (with no bend, no twist).  As described earlier
in Section~\ref{sec:Background}, experimental observations show that
such fibers are characterized by a well-defined orbital motion that can
be separated into one of several distinct orbit classes according to a
fiber flexibility parameter $\chi$ that captures the ratio of fiber
bending force to hydrodynamic drag.  This section aims to investigate
the full range of these two-dimensional orbital motions.

In all simulations, we use the numerical parameters listed in
Tables~\ref{Table:RigidFiberParameters}
and~\ref{Table:FlexibleFiberParameters}. Since the fiber motion is 
confined to the $xy$-plane, we significantly reduce the execution time 
of a simulation by shrinking the domain depth $H_z$, which 
allows us to run $100+$ simulations in a reasonable timeframe.
Note that these results are virtually identical to simulations 
using a larger domain ($H_z=2$), which we confirmed through 
numerous computational experiments.
In all simulations, we choose physical parameters that are
consistent with natural (unbeaten) kraft pulp fibers, taking a fiber
length of $0.1-0.3$~cm and flexural rigidity of $0.001-0.07\, \text{g
  cm}^3/\text{s}^2$ \cite{Tamdoo1981,Tamdoo1982}.  Because fibers in our
numerical simulations have diameter that is proportional to the
effective thickness $w$, our simulated fibers are actually thicker than
a natural pulp fiber.  For example, we use a delta function
regularization corresponding to $w \approx 80~\mu$m, whereas a natural
pulp fiber has a diameter between $20$--$80~\mu$m.  Since the precise
dependence of the simulated fiber diameter on $w$ is unknown, we appeal
to the work of Bringley and Peskin~\cite{BringleyPeskin2008} where they
observed that a one-dimensional array of rigid IB points has an
effective numerical thickness of $D \approx 2 w$. Although these results
may not be strictly applicable in the present setting, this
approximation is sufficient for our purposes.  Any remaining discrepancy
in the fiber diameter can then be accommodated for by adjusting the
value of fiber drag force (see $F_d$ from
equation~\eqref{eq:DragForce}).

\begin{table}[ht!bp]\centering\small
  \caption{Numerical and physical parameter values used in rigid fiber
    simulations.}   
  \begin{tabular}{l|ccc}\toprule
    Parameter & Symbol & Value \\
    \midrule
    Size of fluid domain $\Omega$ & $H_x \times H_y \times H_z$ & $2 \times \half \times 16h$ & cm\\
    Number of fluid grid points & $N_x \times N_y \times N_z$ & $256 \times 64 \times 16$ & \\
    Fluid mesh width & $h$ & $1/128$ & cm\\
    Fluid density & $\rho$ & $1.0$ & g/cm$^3$\\
    Fluid viscosity & $\mu$ & $10.0$ & g/(cm$\cdot$s)\\
    Speed of moving plates & $U_{\text{top}}=U_{\text{bot}}$ & $8$ & cm/s \\
    Shear rate & $G$ & $32$ & s$^{-1}$\\
    Time step & $\dt$ & $1\EE{-5}$ & s \\
    Fiber length & $L$ & $0.3$ & cm\\
    Fiber mesh width & $\GIBds$ & $L/120$ & cm\\
    Bending and twisting modulus (EI)& $a_1=a_2=a_3$ & $0.7$ & $\text{dyne} \cdot \text{cm}^2$\\
    Shear and stretch modulus & $b_1=b_2=b_3$ & $540$ & $\text{dyne} \cdot \text{cm}^2$ \\
    Fiber effective thickness & $w$ & $0.0078125$ & cm\\
    Intrinsic twist vector & $\brac{\kappa_1,~\kappa_2,~\tau}$ & $\brac{0,~0,~0}$ & \\
    Fiber length perturbation & $\epsilon_0$ & $0.001$ &\\
    Support of delta kernel & $C$ & $4$ & \\
    \bottomrule
  \end{tabular}
  \label{Table:RigidFiberParameters}
\end{table}

\begin{sidewaysfigure}[!tbp]
  \begin{center}
    \subfigure[Rigid Orbit ($\chi = 0.19$, $EI = 7.0\EE{-1}$, $L = 0.3$)]{
      \includegraphics[width=0.7\textwidth]{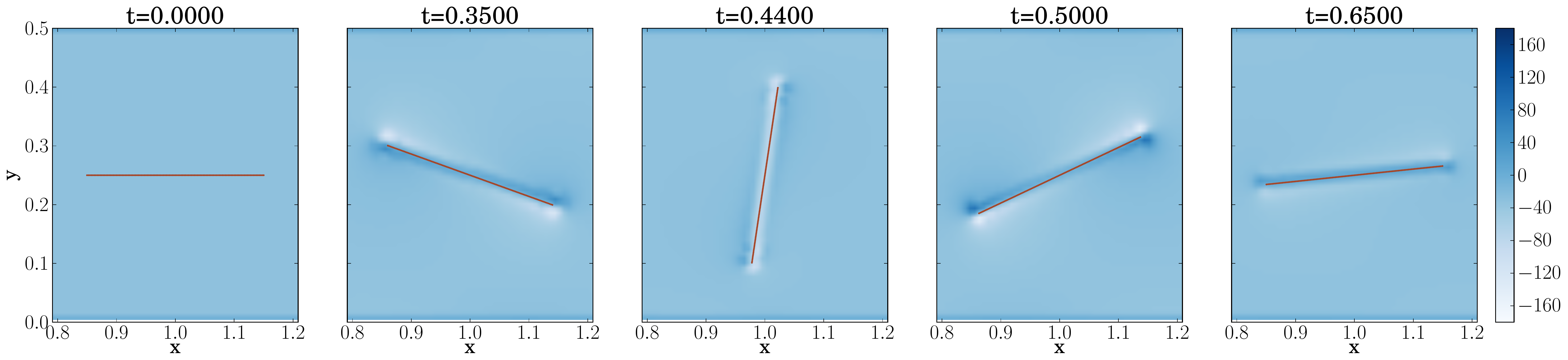}
      \label{fig:RigidOrbit}}
    
    \subfigure[Springy Orbit ($\chi = 4.49$, $EI = 2.5\EE{-2}$, $L \approx 0.282$)]{
      \includegraphics[width=0.7\textwidth]{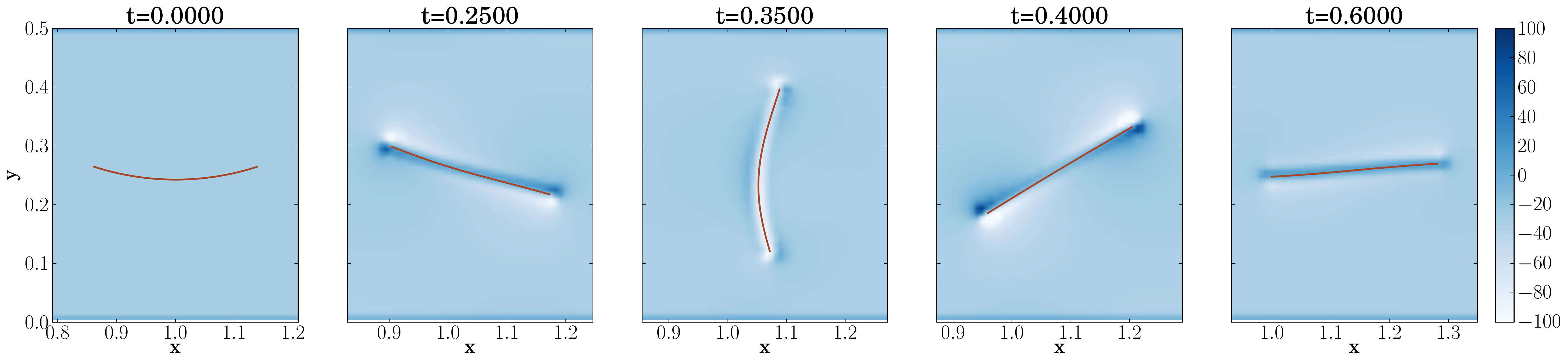}
      \label{fig:SpringyOrbit}}
    
    \subfigure[Snake Orbit ($\chi = 37.38$, $EI = 3.0\EE{-3}$, $L \approx 0.282$)]{
      \includegraphics[width=0.7\textwidth]{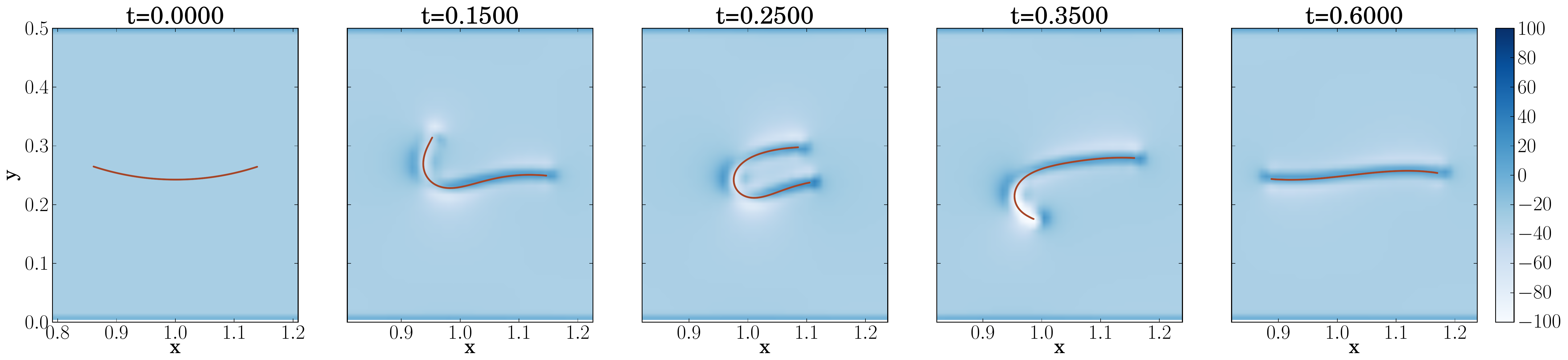}
      \label{fig:SnakeOrbit}}
    \caption{Snapshots of fiber position and fluid vorticity in the
      $xy$-plane for a half-rotation in a rigid, springy and snake
      orbit.  Parameter values are listed in
      Tables~\ref{Table:RigidFiberParameters} and
      \ref{Table:FlexibleFiberParameters}.}
    \label{fig:AllOrbits} 
  \end{center}
\end{sidewaysfigure}

\begin{sidewaysfigure}[!tbp]
  \begin{center}
      \subfigure[S~Orbit ($\chi = 45.00$, $EI = 3.0\EE{-3}$, $L = 0.3$)]{
       \includegraphics[width=0.7\textwidth]{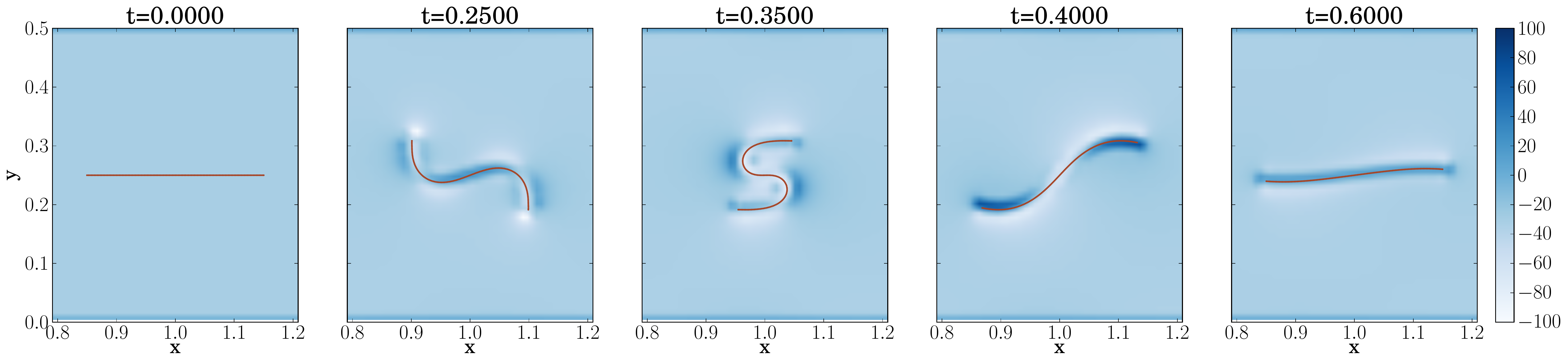}
      \label{fig:SOrbit}}
    \subfigure[Complex Orbit ($\chi = 119.06$, $EI = 1.0\EE{-3}$, and $L \approx 0.251$)]{
       \includegraphics[width=0.7\textwidth]{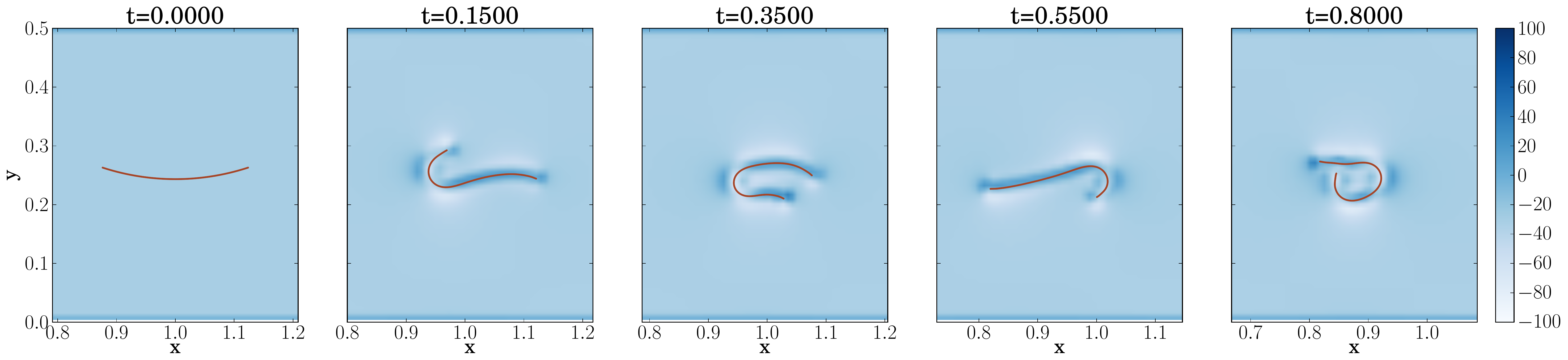}
      \label{fig:ComplexOrbit}}
    \subfigure[Coiled Orbit ($\chi = 1.125\EE{5}$, $EI = 1.0\EE{-4}$, and $L = 0.5$)]{
       \includegraphics[width=0.75\textwidth]{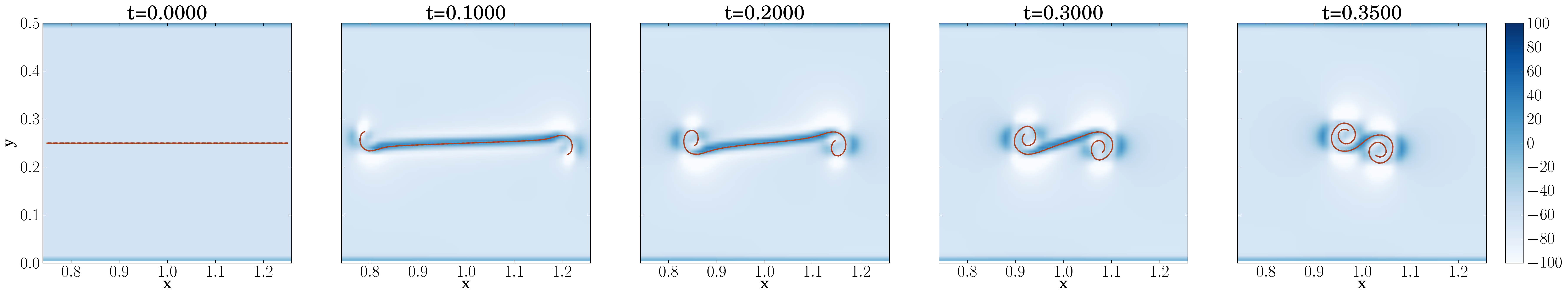}
      \label{fig:CoiledOrbit}}
    \caption{Snapshots of fiber position and fluid vorticity in the
      $xy$-plane for an S~turn, complex and coiled orbit.  Parameter
      values are listed in Tables~\ref{Table:RigidFiberParameters} and
      \ref{Table:FlexibleFiberParameters}.}
    \label{fig:AllOrbits2} 
  \end{center}
\end{sidewaysfigure}

In Figures~\ref{fig:AllOrbits} and~\ref{fig:AllOrbits2}, we display
snapshots of the dynamics of a fiber with initial configuration lying in
the $xy$-plane and for six values of the dimensionless flexibility
parameter $\chi$ between $0.19$ and $1.125\EE{5}$.  As expected, the
simulations exhibit a range of different orbital motions that transition
between the various orbit classes (rigid, springy, flexible, complex,
coiled) as the flexibility increases.  We also note that within the
intermediate range of $\chi$ values, we observe both S~turns and snake
turns depending on the symmetry of the initial fiber configuration.
Despite being very rare in actual fiber suspensions, S~turns turn
out to be remarkably stable in our idealized setting with a planar shear
flow; indeed, it is only when asymmetry is introduced in the fiber
through (for example) the initial shape or a length-dependent stiffness
that snake turns are observed instead of S~turns.  These results are
consistent with those of Mason and
co-workers~\cite{Arlov1958,Forgacs1959} who observed that S~turns
required a high degree of symmetry that is rarely achieved in
experiments.  For the largest value of $\chi=1.125\EE{5}$ in
Figure~\ref{fig:AllOrbits2}\subref{fig:CoiledOrbit} we observe a coiled
orbit with self-entanglement, and although this type of behaviour is not
pertinent to pulp fibers, Forgacs and Mason~\cite{Forgacs1959} did
observe such coiling with thread-like synthetic fibers.  Eventually,
this fiber forms a complex writhing bundle as the fiber undergoes
self-contact, but because our model doesn't incorporate any contact
(fiber-on-fiber) forces we make no claim that these results correspond
to physically accurate coiling dynamics.

\begin{table}[!tbp]\centering\small
  \caption{Parameter modifications for the flexible fiber simulations in
    Figures~\ref{fig:AllOrbits} and~\ref{fig:AllOrbits2}.  Only those
    parameters that have changed relative to values indicated in
    Table~\ref{Table:RigidFiberParameters} are shown here.}   
  \begin{tabular}{ccc}\toprule
    Orbit Class & Configuration & Parameters \\
    \midrule
    Springy & 2 & $r_0 =0.45$, $\alpha_b=0.4$, $\alpha_e=0.6$, $EI=2.5\EE{-2}$, \\
            & & $\GIBds\approx 1.25\EE{-3}$, $L\approx 0.282$\\\\
    S~turn  & 1 & $EI=3.0\EE{-3}$\\\\
            & & \\
    Snake turn & 2 & $r_0 =0.45$, $\alpha_b=0.4$, $\alpha_e=0.6$, $EI=3.0\EE{-3}$, \\
            & & $\GIBds\approx 1.25\EE{-3}$, $L\approx 0.282$\\\\
    Complex & 2 & $r_0 =0.4$, $\alpha_b=0.4$, $\alpha_e=0.6$, $\mu = 15$, $EI=1.0\EE{-3}$, \\
            & & $\GIBds\approx 1.25\EE{-3}$, $L\approx 0.251$\\\\
    Coiled  & 1 & $G = 64,$ $\mu = 90$, $EI=1.0\EE{-4}$, $L=0.5$\\
    \bottomrule
  \end{tabular}
  \label{Table:FlexibleFiberParameters}
\end{table}
    
When the initial fiber configuration is rotated into the $xz$-plane, the
resulting dynamics are non-planar but still follow orbits qualitatively
similar to those derived by Jeffery~\cite{Jeffery1922}.  Examples of
these non-planar orbits are given in the first author's doctoral
thesis~\cite{Wiens2014}, which show that the flexible fiber undergoes a
motion consisting of a rotations in the $xy$-plane superimposed on a
rocking motion back and forth about the $z$-axis in the $xz$-plane.

We next explore in more detail the dependence of the fiber orbit class
on the dimensionless flexibility parameter $\chi$.  To this end, we
perform a much larger series of simulations with varying fiber length
($L=0.1$--$0.3$~cm), diameter ($D \approx 156$--$312~\mu$m), flexural
rigidity ($EI = 0.001$--$0.1~\text{dyne} \cdot \text{cm}^2$), shear rate
($G = 20$--$120$~s$^{-1}$) and viscosity ($\mu =
0.07$--$100.0$~g/(cm$\cdot$s)) corresponding to Reynolds numbers lying in
the range $0.0027$--$23.9$.  For each simulation, we assign the fiber
dynamics to one of the four orbit classes \orbiti--\orbitiv\ by
calculating the {\it total fiber curvature}
\begin{align*}
  \lambda(t) = \int_0^L \left|\pd{\bs{D}^3}{s}(s,t)\right| \,ds,
\end{align*}
and using the maximum curvature over a half-rotation $t_0 \leq t \leq
t_1$ to apply the following criteria:
\begin{itemize}
  \itemsep.5em 
\item \orbiti: The orbit is rigid if $\displaystyle \max_{t_0 \leq t \leq
    t_1} \lambda(t)< 0.4$.
\item \orbitii: The orbit is springy if $\displaystyle 0.4 \leq
  \max_{t_0 \leq t \leq t_1} \lambda(t)< 3.7$.
\item \orbitiii: The orbit is an S~or snake turn if $\displaystyle 3.7
  \leq \max_{t_0 \leq t \leq t_1} \lambda(t)$ and $\lambda(t_1)< 2.5$.
\item \orbitiv: The orbit is complex if $\displaystyle 3.7 \leq
  \max_{t_0 \leq t \leq t_1} \lambda(t)$ and $\displaystyle 2.5 \geq
  \lambda(t_1)$.
\end{itemize}
Note that S/snake turns and complex rotations have the same range of
maximum curvature, and that we use the fiber curvature $\lambda(t_1)$ at
the end of the half-rotation to determine whether or not the fiber has
straightened out.

\begin{figure}[!tbp]
  \begin{center}
    \subfigure[]{\includegraphics[width=0.45\textwidth]{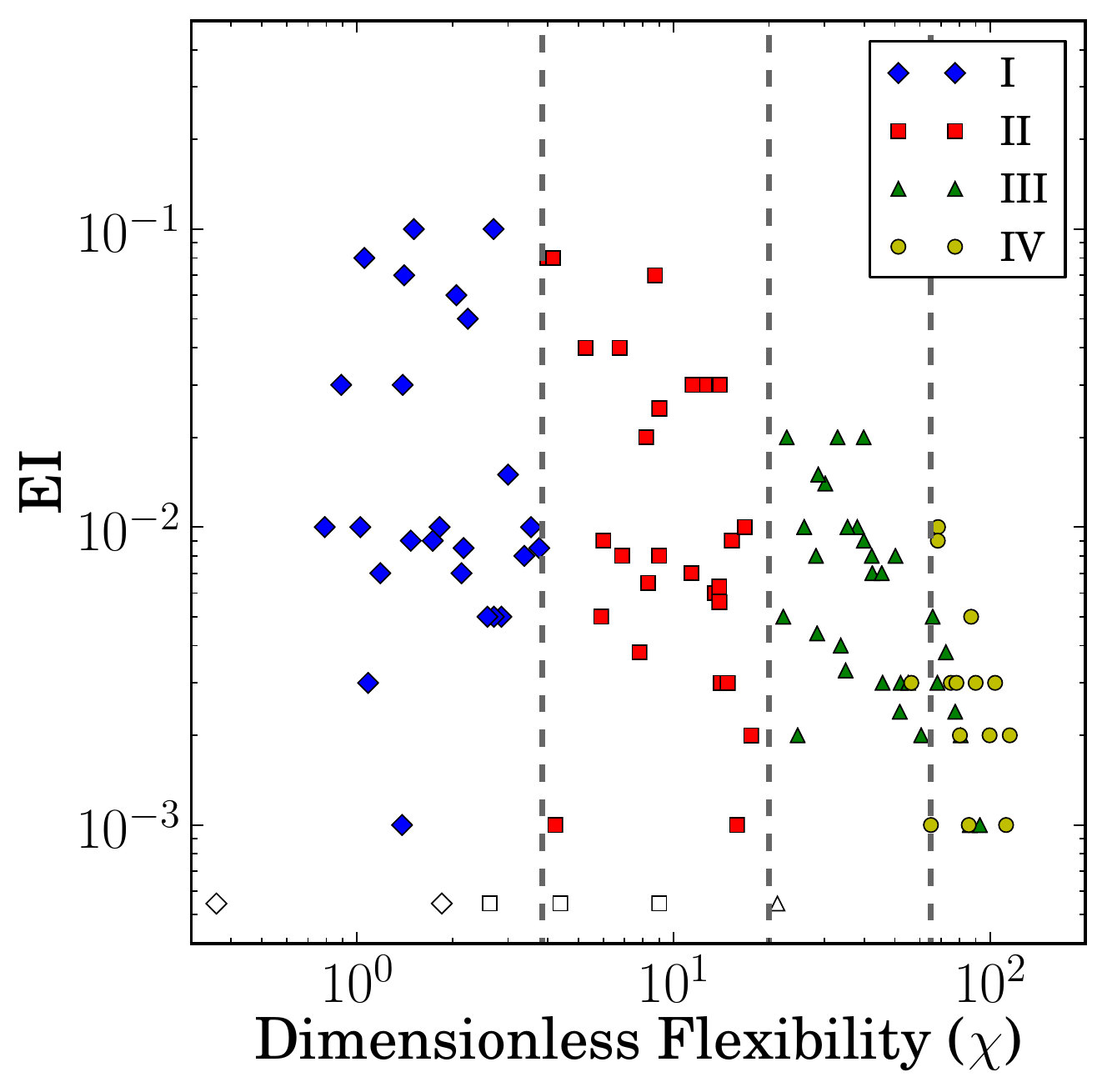} 
      \label{fig:FiberType:EIVsChi}}
    \qquad 
    \subfigure[]{\includegraphics[width=0.45\textwidth]{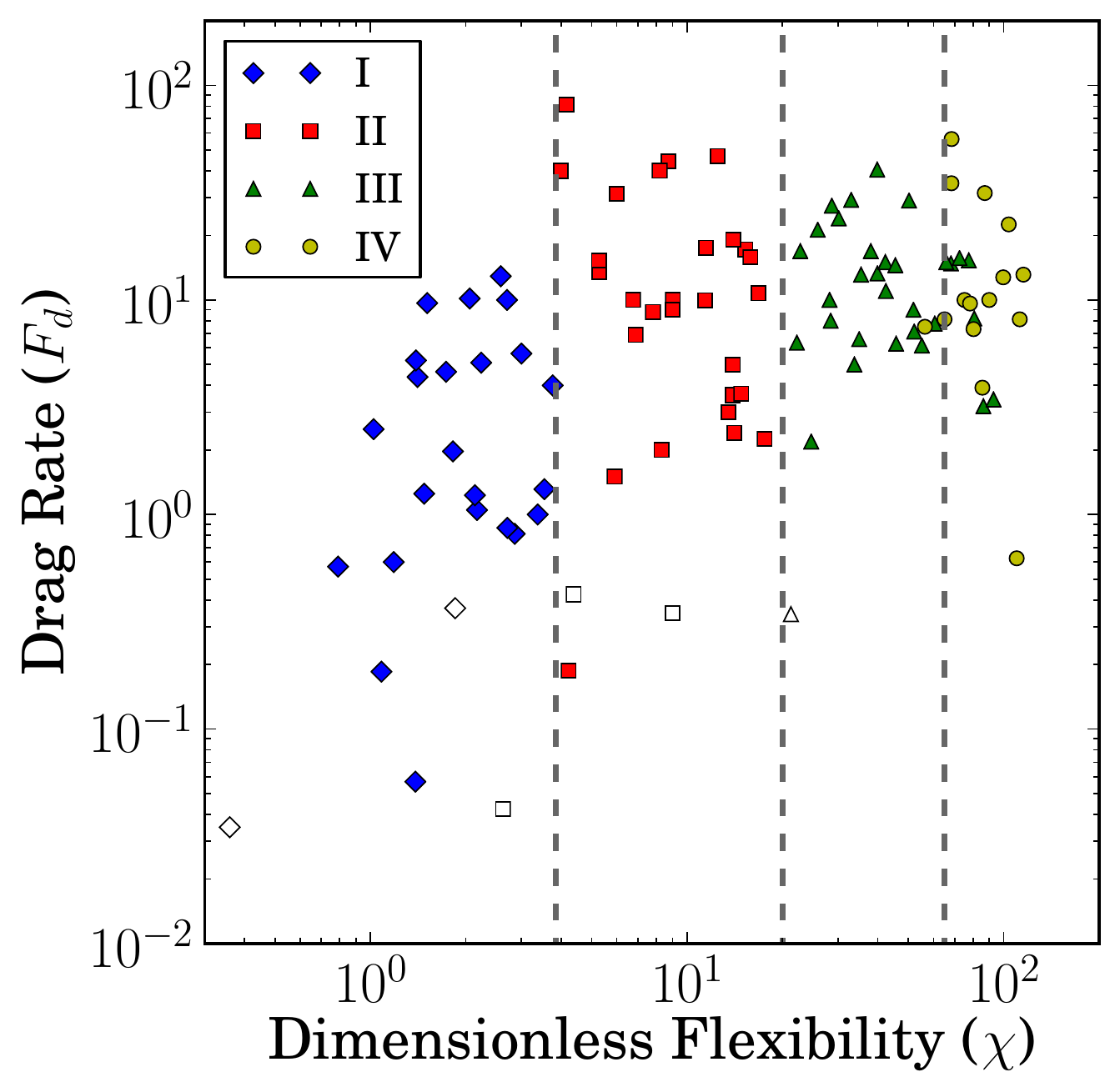} 
      \label{fig:FiberType:FdVsChi}}
    \caption{Summary of all simulations showing the relationship between
      orbit class and different values of the dimensionless flexibility
      $\chi$, flexural rigidity $EI$ and drag rate $F_d$. Open markers denote the experimental data shown in 
              Table~\ref{Table:ExperimentalOrbitClass} where $E=3$ GPa.}
    \label{fig:FiberType} 
  \end{center}
\end{figure}

Simulations are depicted graphically in Figure~\ref{fig:FiberType} in
terms of two plots of flexural rigidity $EI$ and drag force $F_d$ versus
dimensionless flexibility $\chi$.  Each point on the plot corresponds to
a simulation using a specific choice of physical parameters, and the
point type is assigned based on the orbit classification criteria above.
From these two plots, it is evident that there is a clear division of
orbits into classes \orbiti, \orbitii\ and \orbitiii\ along vertical
divisions that correspond to values of $\chi \cong 3.85$ and $\chi \cong
20.0$.  The boundary between classes \orbitiii~and \orbitiv\ is not as
sharply defined, but can still be assigned to a value of flexibility
$\chi \approx 65.0$.  Based on these observations, we conclude that the
dimensionless flexibility $\chi$ provides a useful measure for
characterizing orbit classes at the lower Reynolds numbers considered
here.

We conclude this section by performing a further comparison of our
numerical simulations with the experiments of Forgacs and
Mason~\cite{Forgacs1959} on dacron fibers in corn syrup.  First of all,
we list the parameters and observed orbit class for several of these
experiments in Table~\ref{Table:ExperimentalOrbitClass}.  Based on
values of $\chi \cdot EI$, we see that this rescaled flexibility
parameter may be used to classify each orbit, assuming that $EI$ is
constant in all experiments.  However, we emphasize that since Forgacs
and Mason did not provide a value for the flexural rigidity ($EI$), we
were unable to determine the value of $\chi$ explicitly.
\begin{table}[ht!bp]\centering\small
  \caption{Experimental results obtained from Forgacs and
    Mason~\cite{Forgacs1959} for synthetic dacron fibers.}
  \begin{tabular}{cccccc}\toprule
    Orbit Class & $\chi \cdot EI$ & $G$ (s$^{-1}$) & $\mu$ (g/(cm$\cdot$s)) & $L$ (cm)& $D$ ($\mu$m) \\
    \midrule
    Rigid       & $1.96\EE{-4}$ & $3.921$ &  $11.4$ & $0.1778$ & $7.8$ \\
    Rigid       & $1.01\EE{-3}$ & $5.143$ &  $91.2$ & $0.1404$ & $7.8$ \\
    Springy     & $1.43\EE{-3}$ & $4.763$ &  $11.4$ & $0.3229$ & $7.8$ \\
    Springy     & $2.39\EE{-3}$ & $5.965$ &  $91.2$ & $0.1778$ & $7.8$ \\
    Springy     & $4.91\EE{-3}$ & $4.879$ &  $91.2$ & $0.2418$ & $7.8$ \\
    Flexible    & $1.16\EE{-2}$ & $4.825$ &  $91.2$ & $0.3229$ & $7.8$ \\
    \bottomrule
  \end{tabular}
  \label{Table:ExperimentalOrbitClass}
\end{table}

Because these experiments were all performed with dacron fibers, we next
explore further the assumption that $EI$ is roughly constant, and also
whether the experimental results are consistent with the division of
orbit classes in our simulations in Figure~\ref{fig:FiberType}.  First
of all, we remark that all experimental data points are consistent with our
simulations if $2.46\EE{-4} < EI < 3.71\EE{-4}$ ($\text{dyne} \cdot
\text{cm}^2$).\ \ 
Unfortunately, the Young's modulus $E$ for dacron is known to vary over
an extremely wide range of $71.5 \text{ MPa} \leq E \leq 22.1 \text{
  GPa}$ between various manufacturers~\cite{WolframAlphaDacron}.
However, the manufacturer of the fibers used by Forgacs and Mason was
identified as E.I. du Pont de Nemours and Co., and we were able to find
a patent filed by this company in 1969~\cite{Cope1969} for several
dacron blends that lists a much tighter range for Young's modulus of $2.0
\text{ GPa} < E < 3.5 \text{ GPa}$. Therefore, the hypothetical $EI$ of these 
synthetic fibers would be between $3.63\EE{-4} < EI < 6.36\EE{-4}$, which is consistent 
with our numerical results! Furthermore, most data points are
still classified correctly when the $EI$ falls outside our consistency range 
($2.46\EE{-4} < EI < 3.71\EE{-4}$). To illustrate, we have plotted the experimental 
data in Figure~\ref{fig:FiberType} using open markers, assuming $E = 3$ GPa (giving an $EI = 5.45\EE{-4}$). 
Here, we observe that all experimental data are classified correctly, except for one troublesome data point.
Therefore, we conclude from these results that our simulations are
in excellent agreement with experimental data.

\subsection{Intrinsically Curved Fibers}
\label{sec:IntrinsicallyCurvedFiber}

We next consider single flexible fibers that have an intrinsic curvature
at equilibrium, a situation that is often encountered for natural fibers
such as wood pulp.  We use the base parameter values in
Table~\ref{Table:RigidFiberParameters} and simulate two cases
corresponding to the modifications listed in
Table~\ref{Table:FlexibleBentFiberParameters}.  In both cases, the fiber
is initialized as a curved segment of a circular arc with intrinsic
twist vector $\brac{\kappa_1,~\kappa_2,~\tau} = \brac{1/r_0,~0,~0}$, which
keeps the initial fiber configuration at equilibrium (that is,
$N^1=N^2=N^3=0$ at $t=0$).

\begin{table}[!tbp]\centering\small
  \caption{Parameter modifications for the flexible fiber simulations in
    Figure~\ref{fig:ProtrudedSTurn} and~\ref{fig:ProtrudedSnakeTurn}.
    Only those parameters that have changed relative to values 
    in Table~\ref{Table:RigidFiberParameters} are shown here.}   
  \begin{tabular}{ccc}\toprule
    Orbit Class & Configuration & Parameters \\
    \midrule
    S~turn     & 3 & $H_z = 2$, $r_0 =0.45$, $\alpha_b=0.4$, $\alpha_e=0.6$, $EI=3.0\EE{-3}$, \\
               & & $\epsilon_0 = 1\EE{-3}$, $\GIBds\approx 1.25\EE{-3}$, $L\approx 0.282$\\
               \\
    Snake turn & 2 & $H_z = 2$, $r_0 =0.45$, $\alpha_b=0.4$, $\alpha_e=0.6$, $EI=3.0\EE{-3}$, \\
               & & $\Theta_{xz} = \pi/16$, $\GIBds\approx 1.25\EE{-3}$, $L\approx 0.282$\\
    \bottomrule
  \end{tabular}
  \label{Table:FlexibleBentFiberParameters}
\end{table}

\begin{sidewaysfigure}[!tbp]
  \begin{center}
    \includegraphics[width=0.31\textwidth]{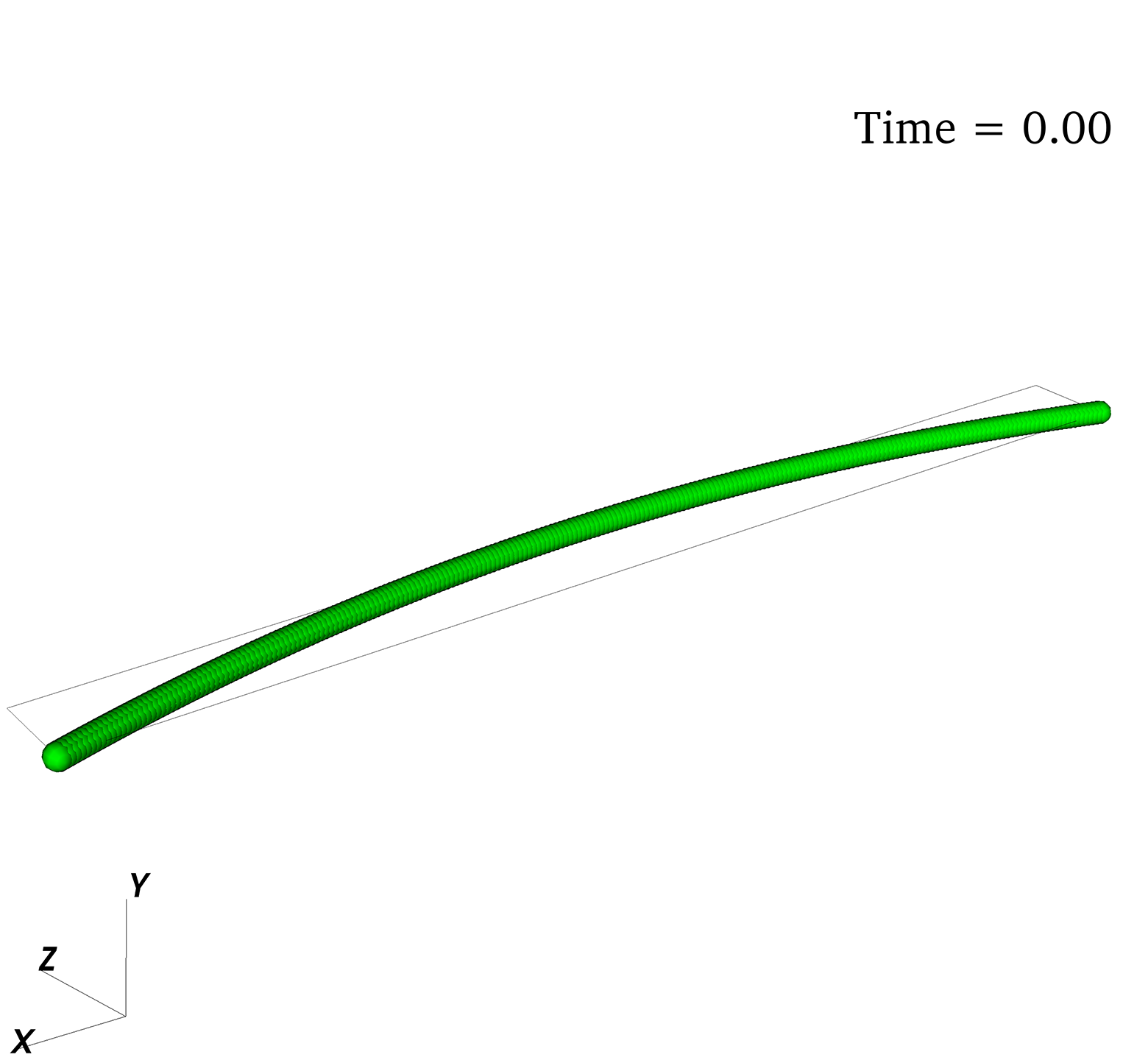} 
    \quad 
    \includegraphics[width=0.31\textwidth]{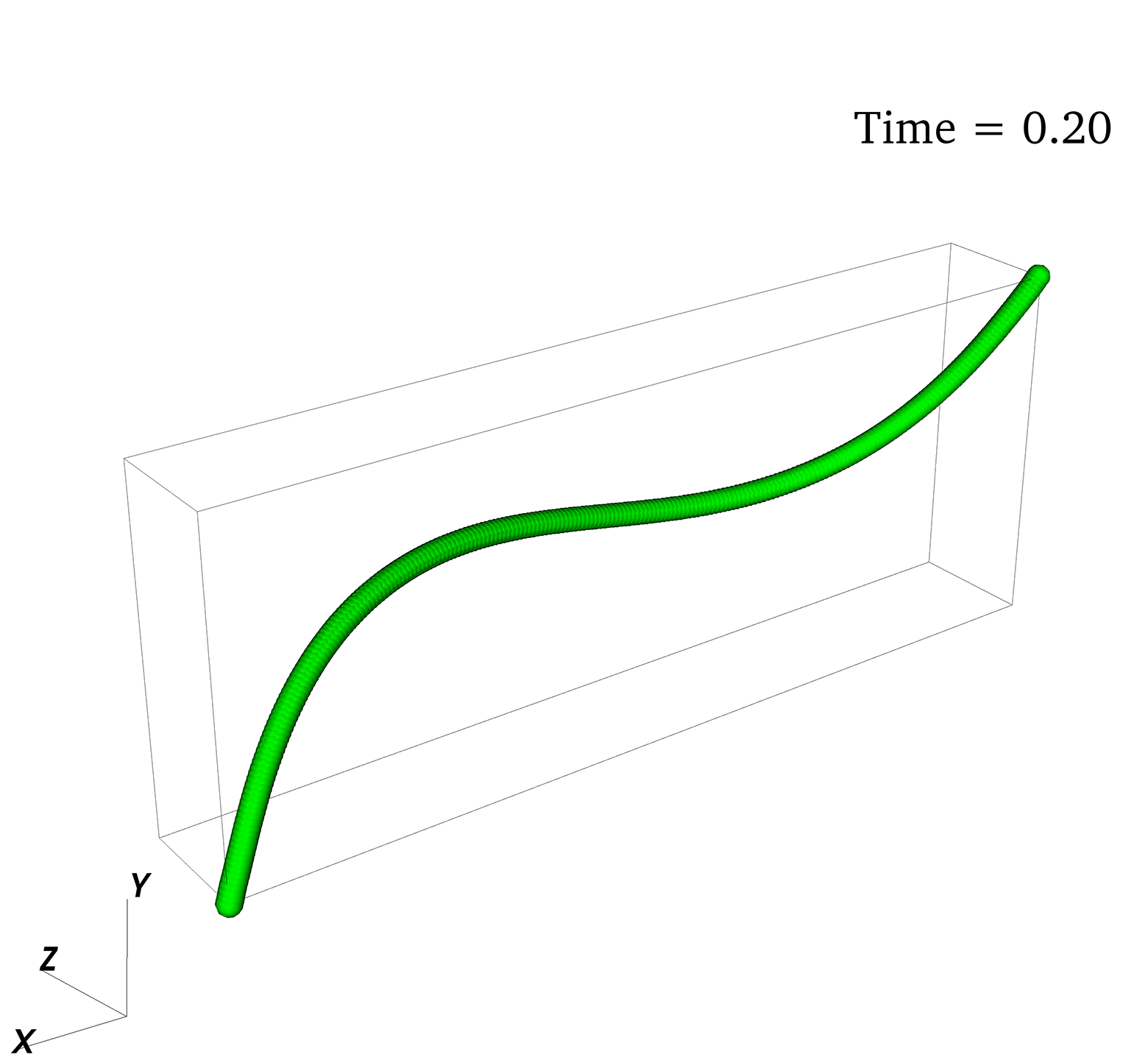} 
    \quad 
    \includegraphics[width=0.31\textwidth]{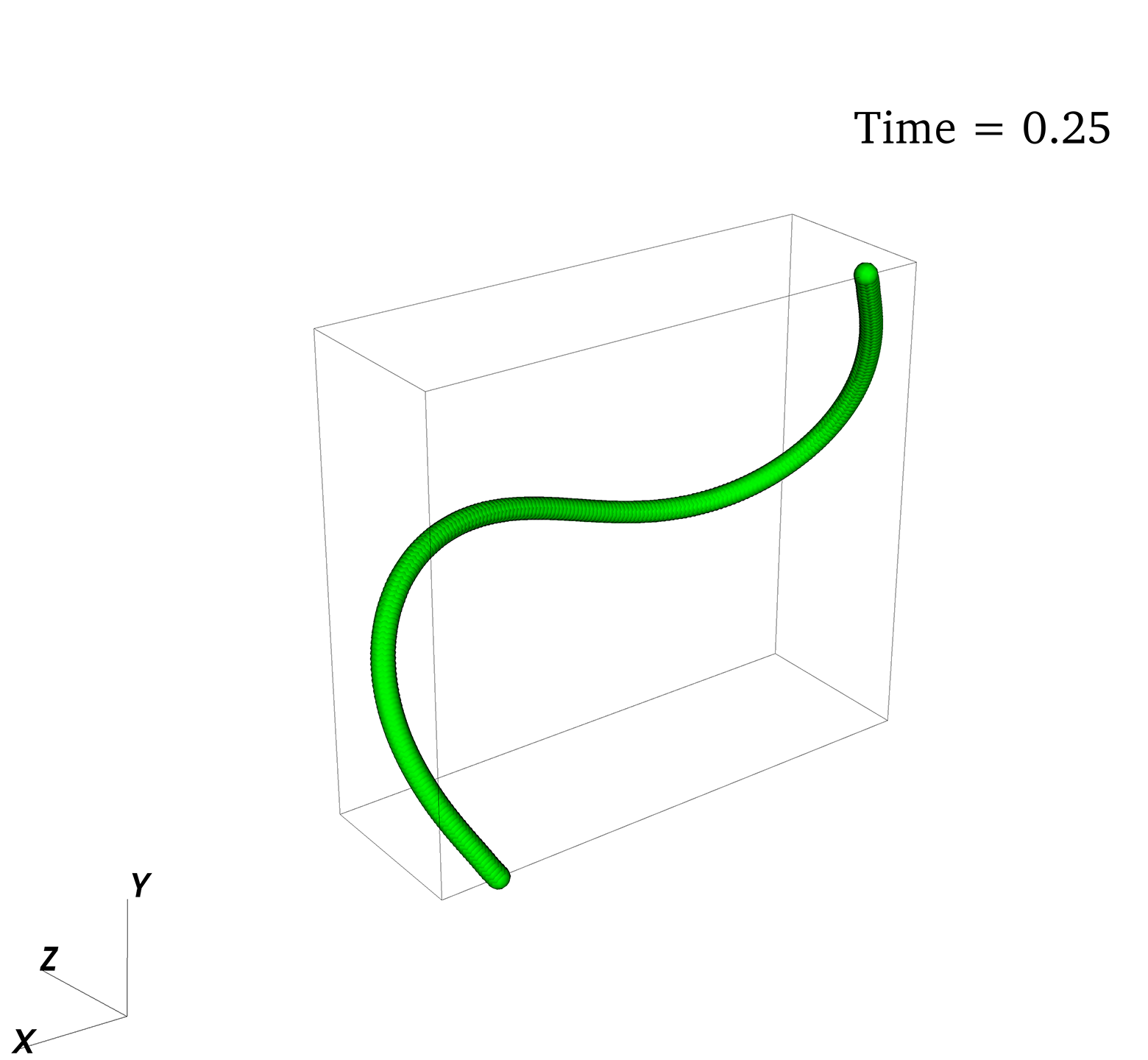} 
    \quad 
    \includegraphics[width=0.31\textwidth]{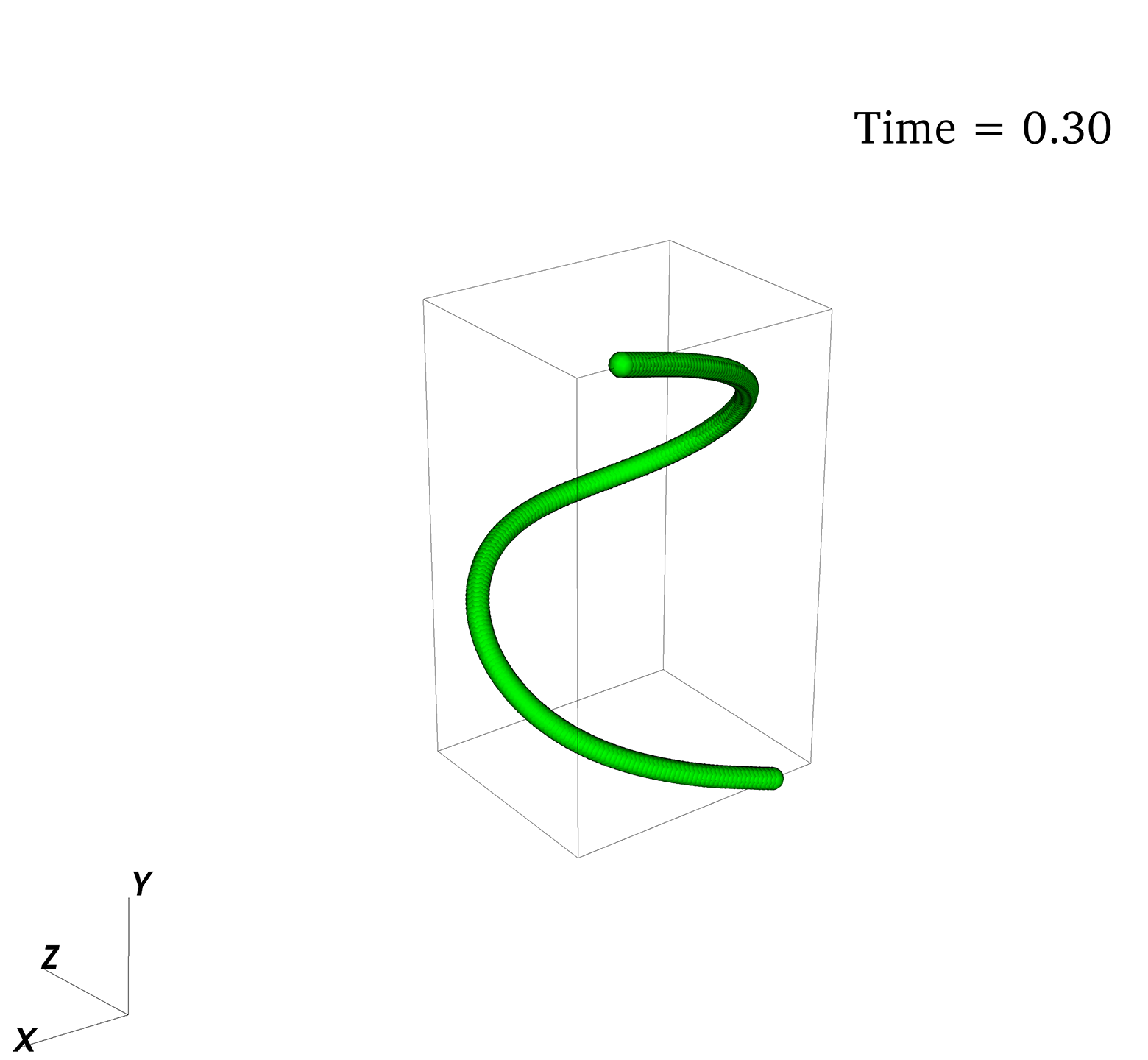} 
    \quad 
    \includegraphics[width=0.31\textwidth]{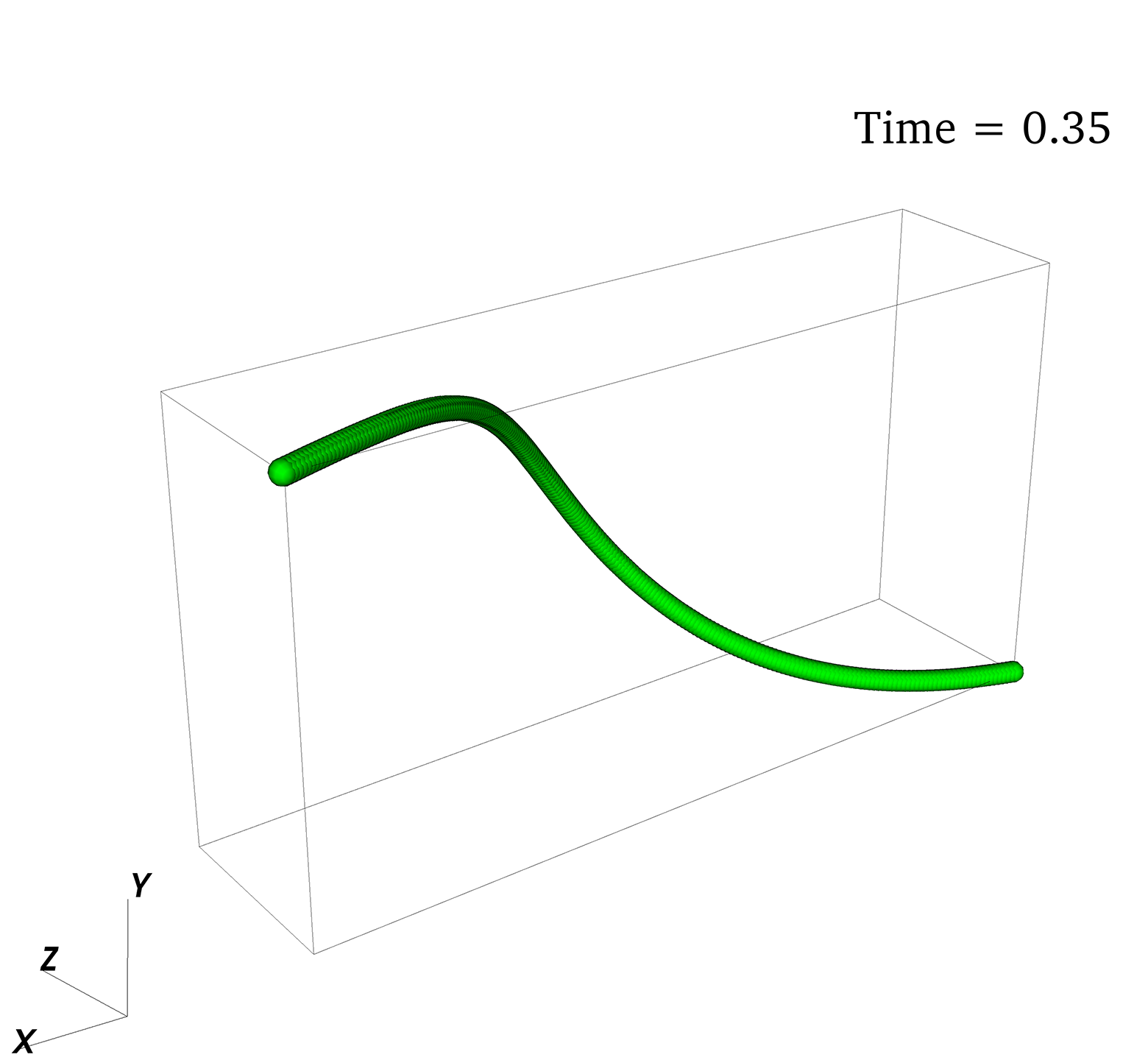} 
    \quad 
    \includegraphics[width=0.31\textwidth]{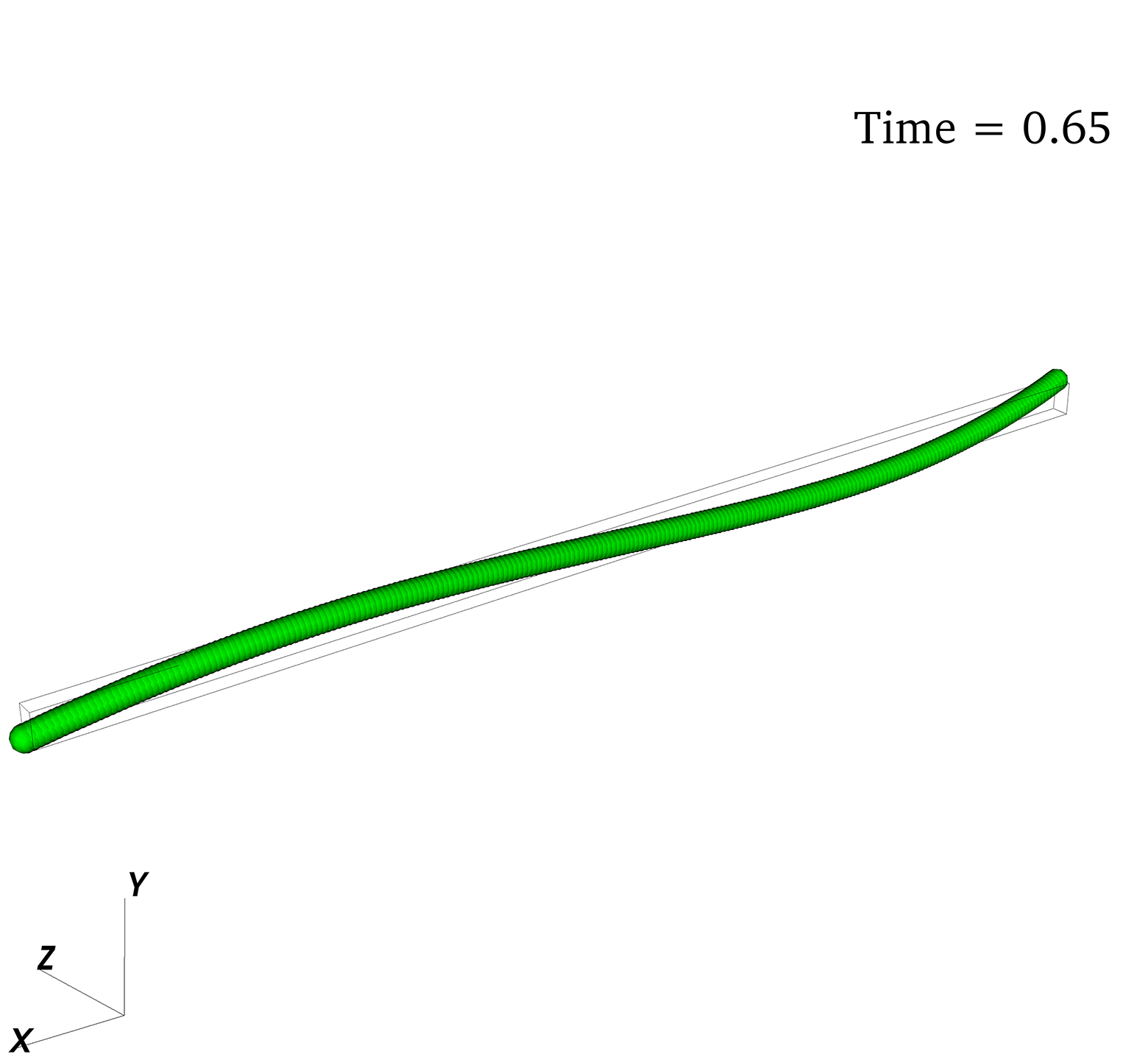} 
    \caption{Snapshots of an S~turn orbit for an intrinsically curved
      fiber with parameters in Tables~\ref{Table:RigidFiberParameters}
      and~\ref{Table:FlexibleBentFiberParameters}.}
    \label{fig:ProtrudedSTurn} 
  \end{center}
\end{sidewaysfigure}

\begin{sidewaysfigure}[!tbp]
  \begin{center}
    \includegraphics[width=0.31\textwidth]{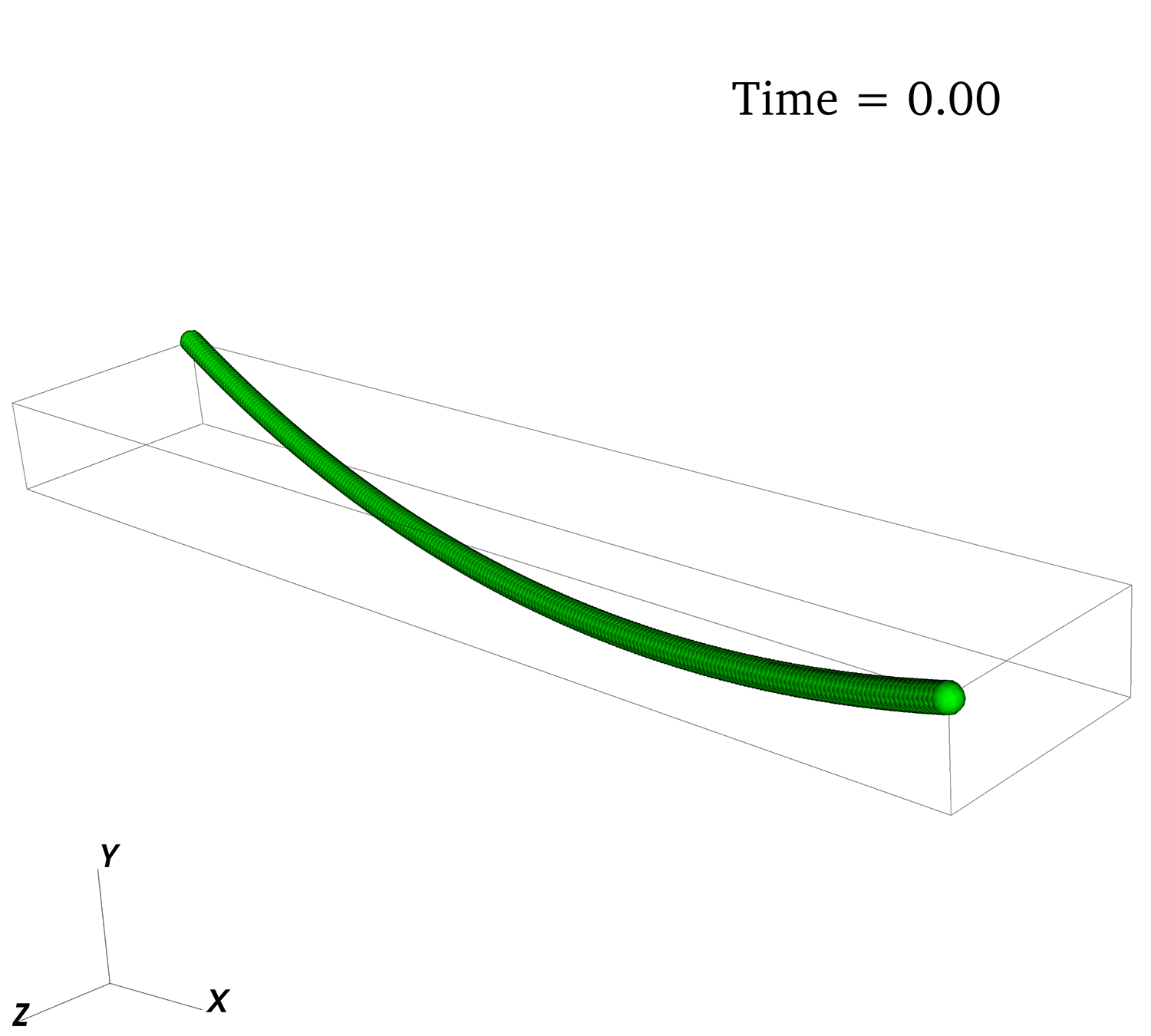} 
    \quad 
    \includegraphics[width=0.31\textwidth]{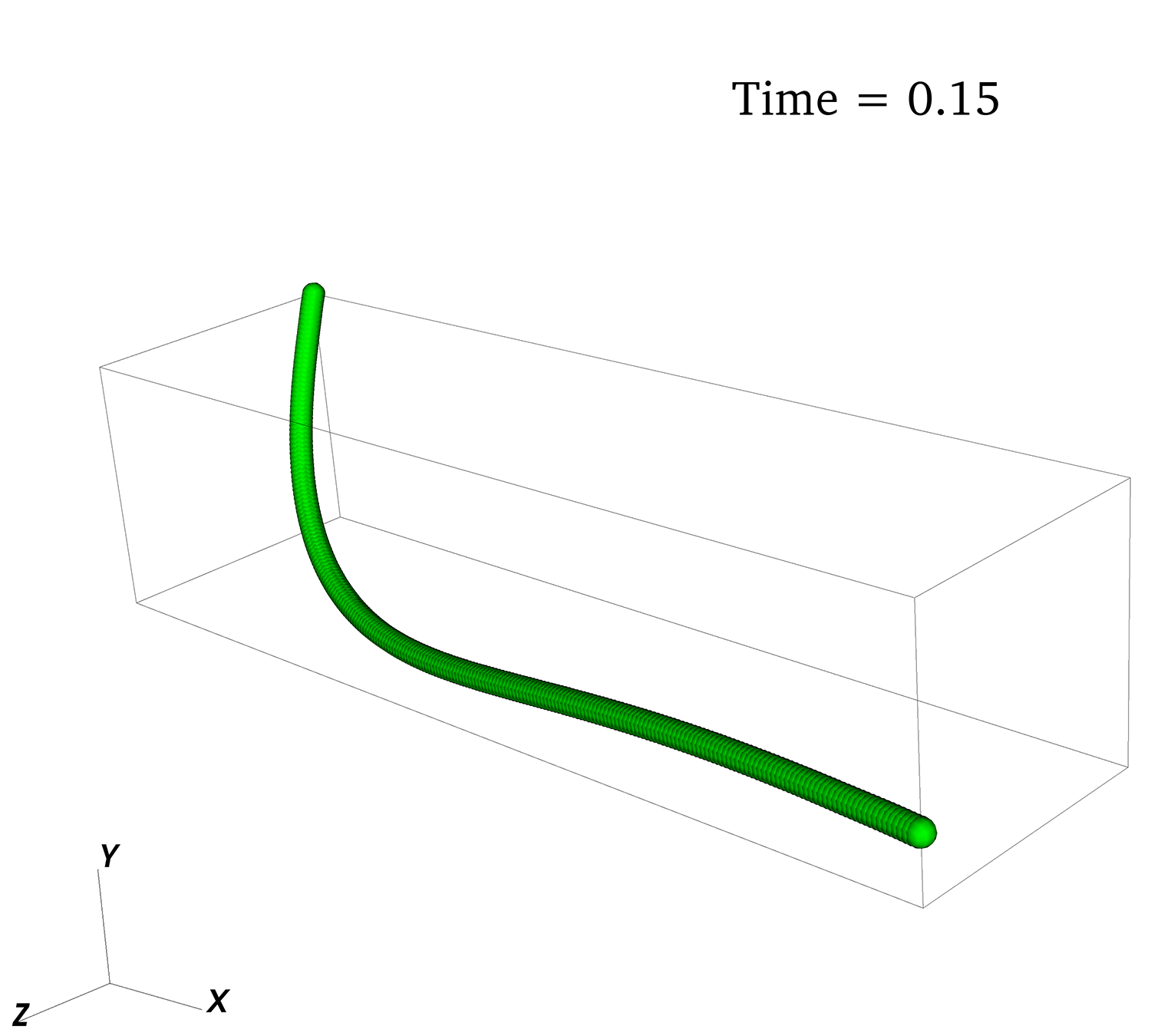} 
    \quad 
    \includegraphics[width=0.31\textwidth]{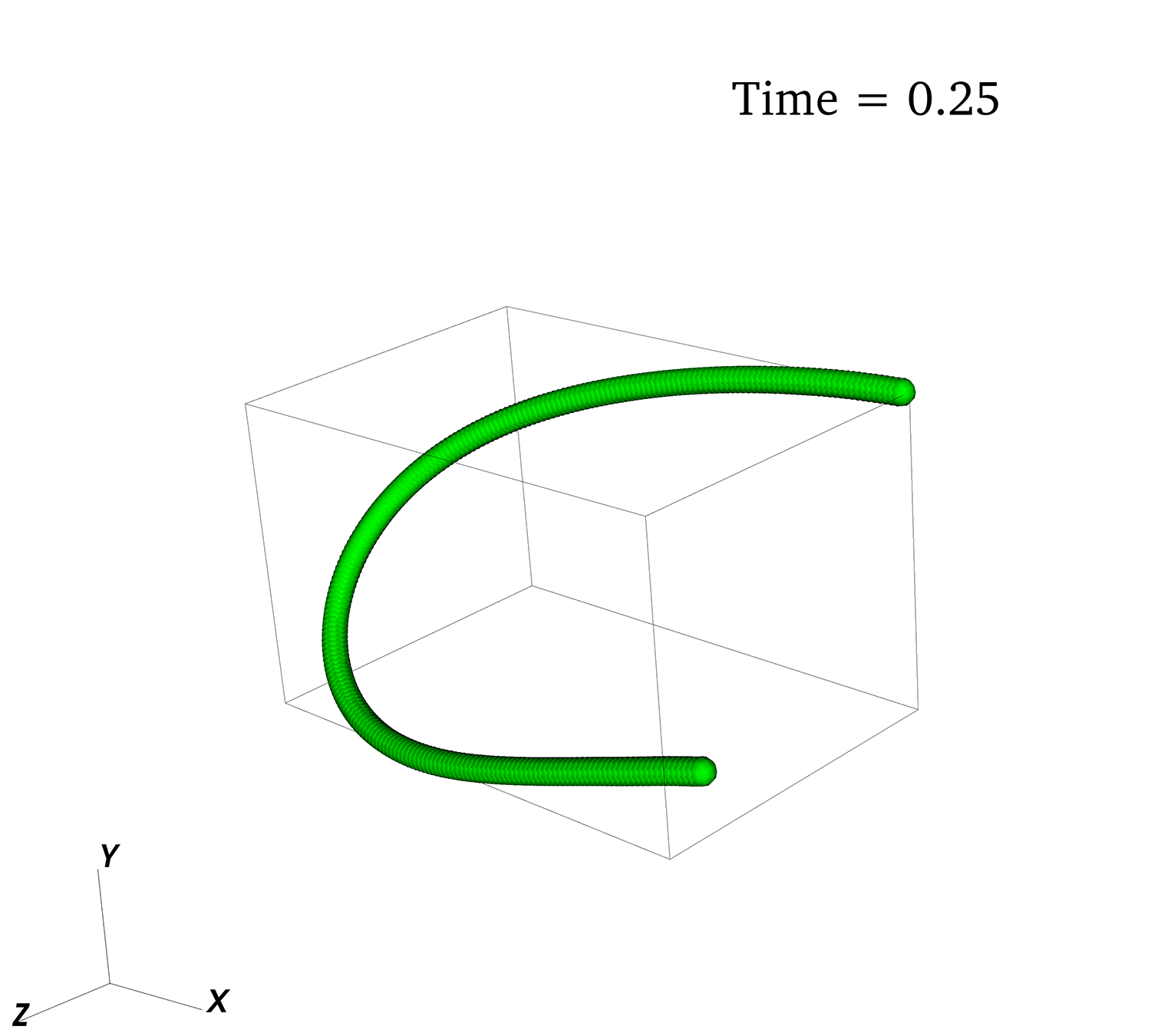} 
    \quad 
    \includegraphics[width=0.31\textwidth]{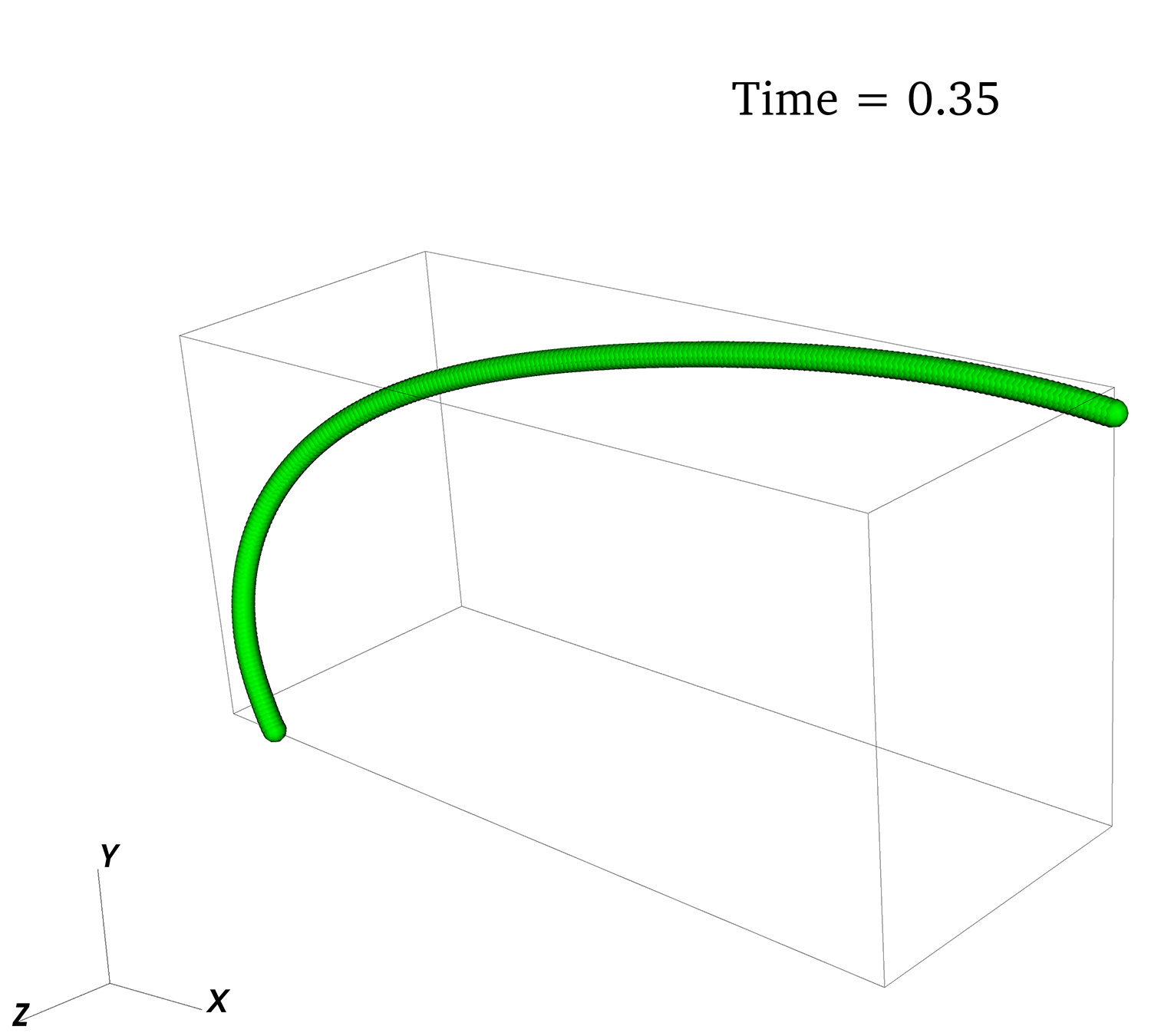} 
    \quad 
    \includegraphics[width=0.31\textwidth]{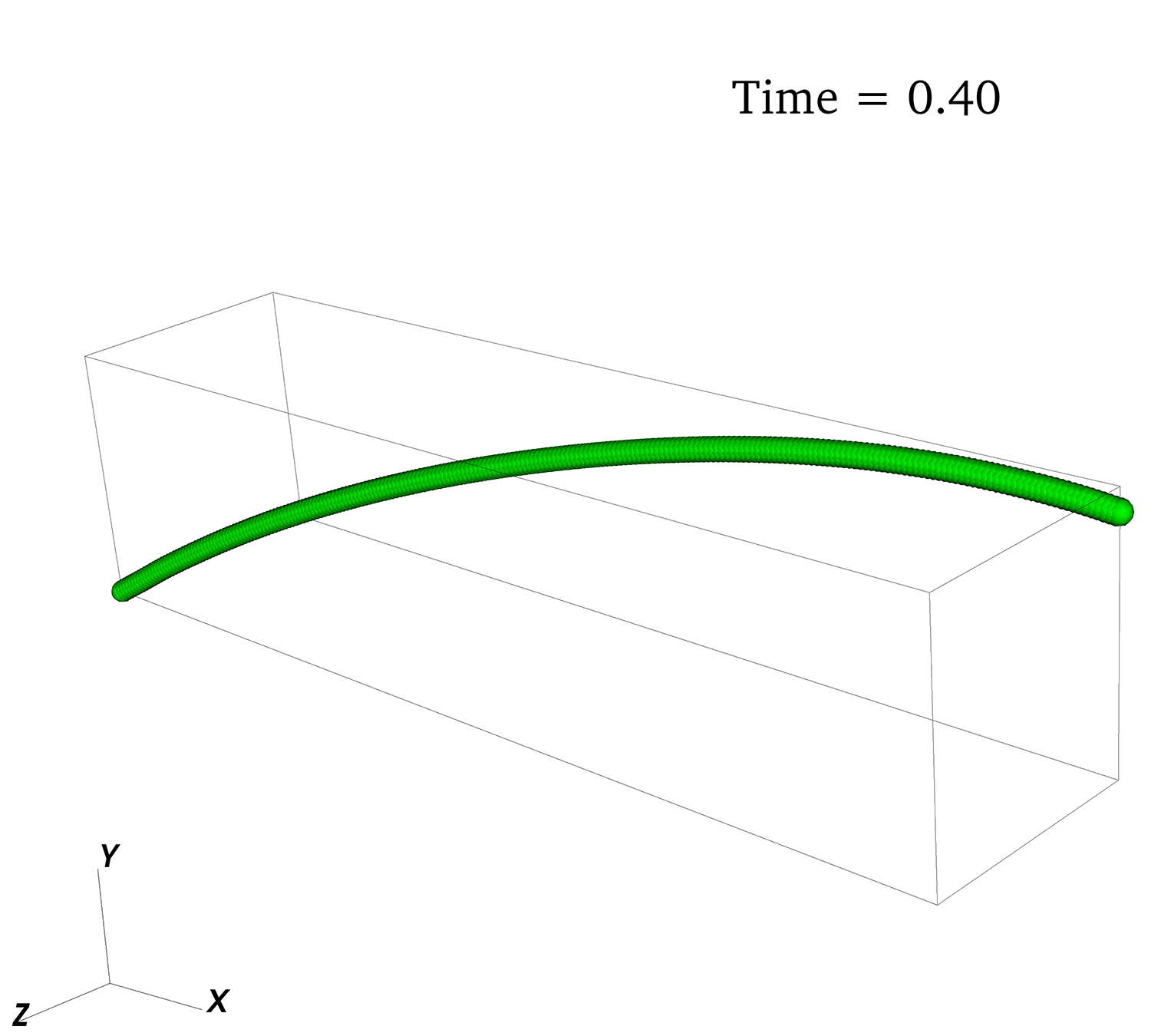} 
    \quad 
    \includegraphics[width=0.31\textwidth]{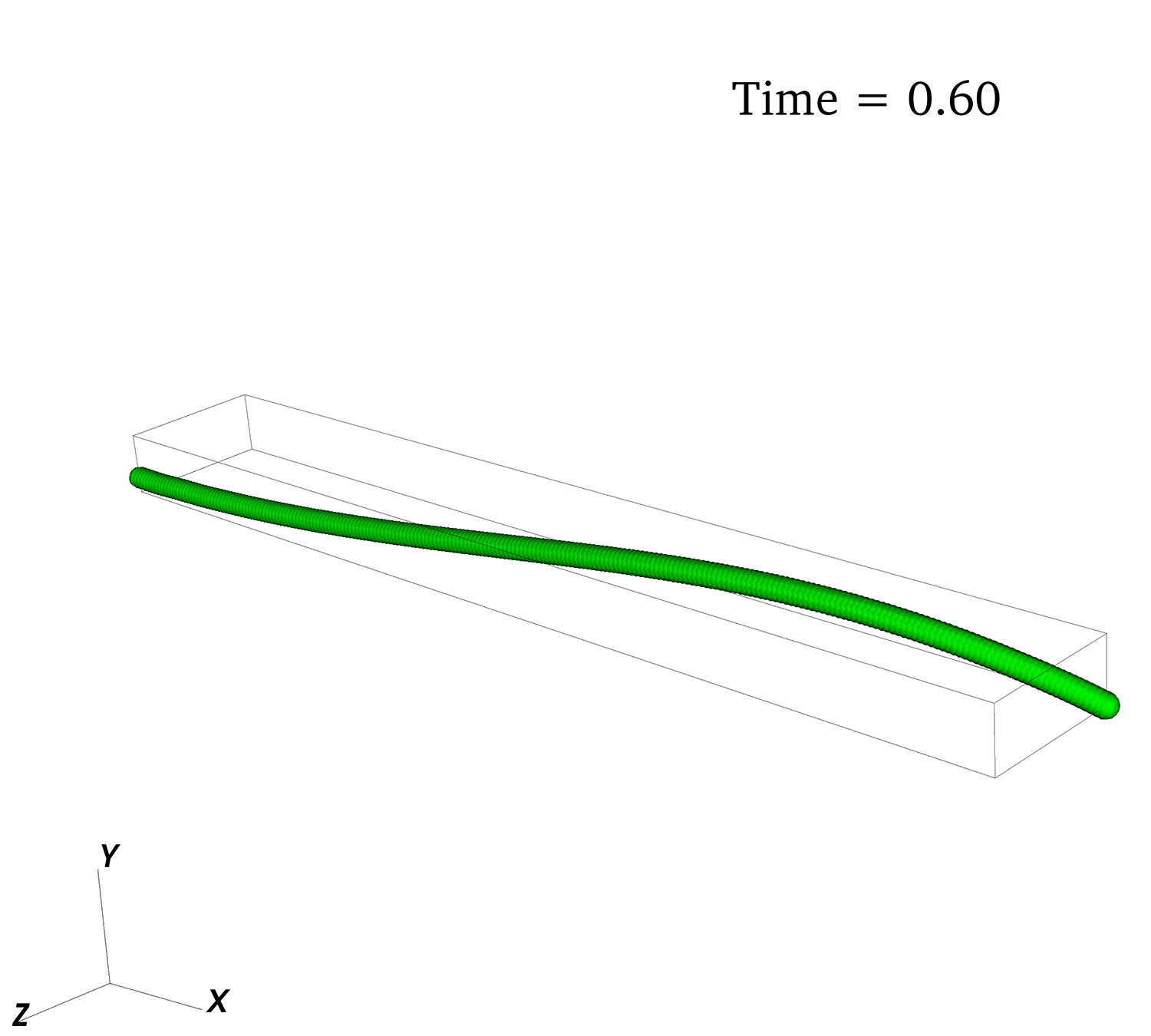} 
    \caption{Snapshots of snake turn for an intrinsically curved fiber
      with parameters in Tables~\ref{Table:RigidFiberParameters}
      and~\ref{Table:FlexibleBentFiberParameters}.}   
    \label{fig:ProtrudedSnakeTurn} 
  \end{center}
\end{sidewaysfigure}

The resulting orbits depicted in Figures~\ref{fig:ProtrudedSTurn}
and~\ref{fig:ProtrudedSnakeTurn} clearly correspond to S- and snake-like
orbits.  The projections of both fibers in the $xy$-plane behave like
the corresponding planar orbits considered in
Section~\ref{sec:IntrinsicallyStraightFiber}, but protrude into the
$xz$-plane.  These simulations reproduce similar orbital dynamics to
those observed in experiments of Arlov et al.~\cite{Arlov1958}.  The
first author's thesis~\cite{Wiens2014} shows additional simulations for
a fiber initially oriented along the $z$-direction and undergoing an
additional axial spin, for which the fiber rotates around the $z$-axis
and slightly straightens out as it rotates into the shear flow.

\subsection{Multiple Flexible Fibers}
\label{sec:MultipleFiberSuspension}

For our last series of simulations, we consider an idealized
representation of a fiber suspension that permits us to employ the
domain tiling techniques described in~\cite{Wiens2013}.  In these
computations, we simulate a $P_x \times 1 \times P_z$ array of fibers
immersed in the fluid domain $\Omega = [0,P_x H_x] \times [0,H_y] \times
[0,P_z H_z]$ using the boundary conditions stated in Section~\ref{sec:InitialConditions}.  The
code runs in parallel on a $P=P_x \times P_z$ array of computer
processors ($P_y=1$) and the fluid domain $\Omega$ is partitioned along
the $x$- and $z$-axes so that one processor labelled $I,K$ is
responsible for each subdomain $\Omega_{I,K}=[(I-1)H_x, IH_x] \times [0,
H_y] \times [(K-1)H_z, KH_z]$, for $I=1,2,\ldots, P_x$ and $K=1,2,\ldots
P_z$.  We have constructed this problem so that it can be used as a weak
scalability test, wherein the local problem size is held fixed as both
the number of processors and global problem size are increased.  It is
important to recognize that our method is in no way restricted to such
idealized arrays of fibers, but rather we have employed this arrangement
here in order to clearly illustrate the parallel scalability of our
algorithm.

Initially, each subdomain $\Omega_{I,K}$ contains a single
intrinsically-curved fiber located at its centroid, with a
randomly-chosen orientation angle and whose initial shape is defined in
the same manner as described earlier for Configuration~3.  The numerical
and physical parameters are as in Table~\ref{Table:RigidFiberParameters}
with the following modifications: $H_x = 0.421875$, $H_y = \half$, $H_z
= 0.3125$, $\dt = 5\EE{-5}$, $r_0 =0.45$, $\alpha_b=0.4$,
$\alpha_e=0.6$, $EI=3.0\EE{-3}$, $\GIBds\approx 1.25\EE{-3}$, $L\approx
0.282$, $U_{\text{top}}=8.5$ and $U_{\text{bot}}=7.5$.  Another
difference from our earlier simulations is that the top and bottom
boundaries that induce the shear flow now move at different speeds (that
is, $U_{\text{top}} \ne U_{\text{bot}}$); consequently, fibers are
transported across subdomain boundaries which provides a nontrivial test
of our algorithm's ability to handle inter-process communication as well
as changes to the IB data stored on each processor over time.

\begin{figure}[!tbp]
  \begin{center}
    \includegraphics[width=0.90\textwidth]{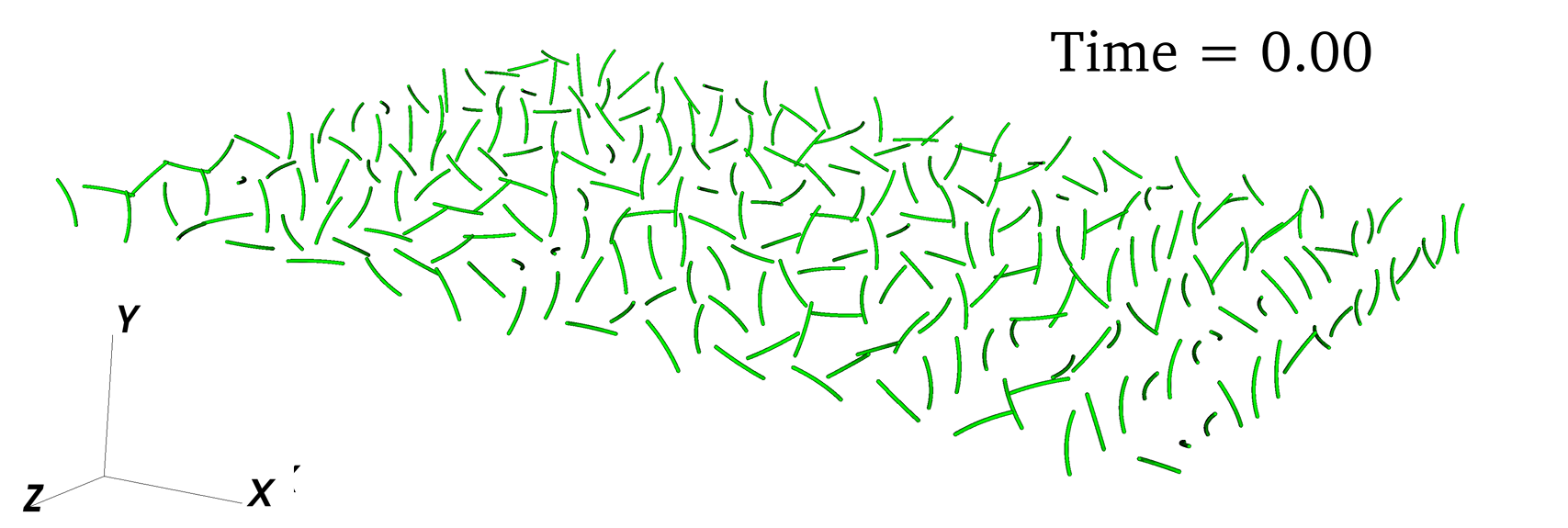} 
    \qquad 
    \includegraphics[width=0.90\textwidth]{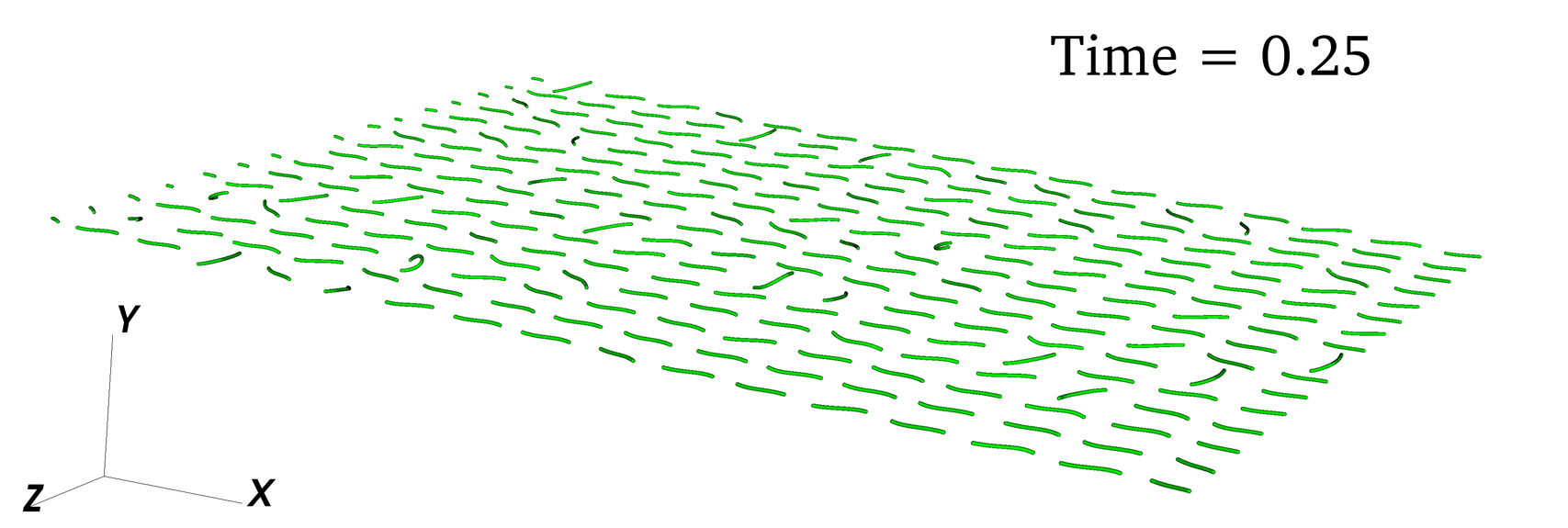} 
    \qquad 
    \includegraphics[width=0.90\textwidth]{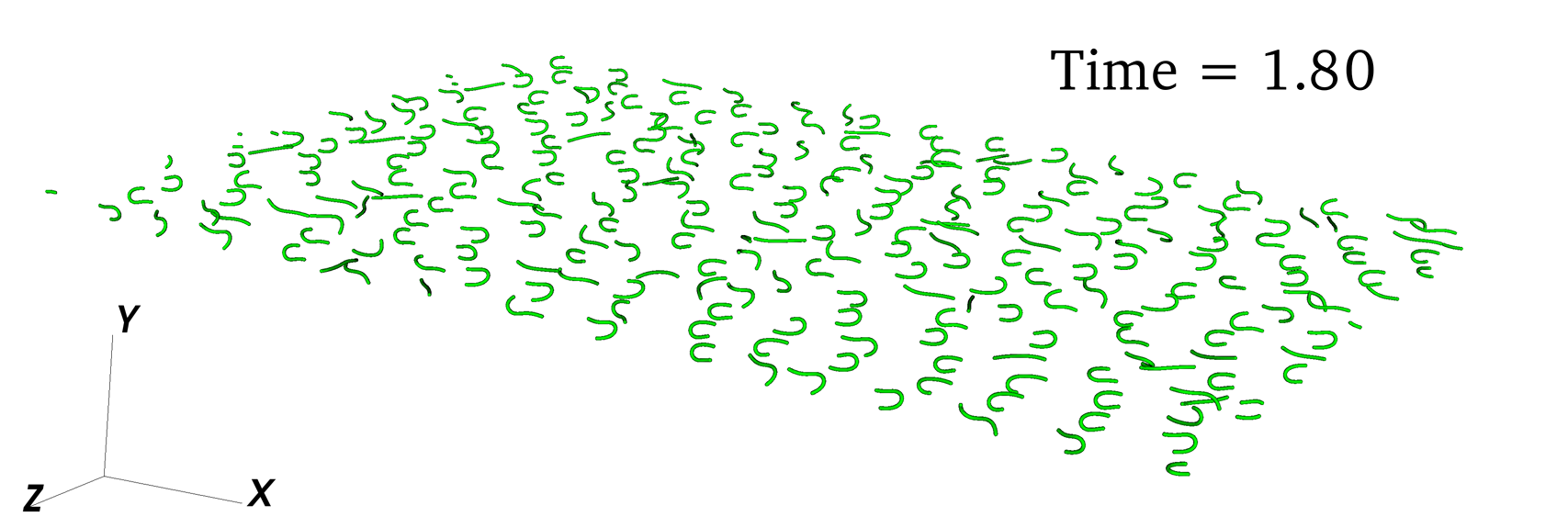}
    \caption{A suspension of $256$ intrinsically-curved fibers ($P_x =
      P_z = 16$) in Configuration $3$.  Parameters are described in
      Section~\ref{sec:MultipleFiberSuspension}.}  
    \label{fig:BigMultiFiberCurvedFlexible} 
  \end{center}
\end{figure}

Figure~\ref{fig:BigMultiFiberCurvedFlexible} presents three snapshots of
the dynamics of a $16 \times 16$ array of fibers at the initial and two
later times.  The image at time $t=0.25$ emphasizes the fact that all
fibers spend the majority of their time aligned horizontally with the
shear flow (i.e., along the $x$-axis) and that only a small proportion
of the fibers at any time instant are rotated out of the shear plane.
As the suspension evolves over time, the fibers are prone to drift and
cluster together, leading to development of more complex behavior such
as is shown in the image at time $t=1.80$.  This last snapshot suggests
that our algorithm is capable of simulating at least the initial phases
fiber flocculation in a suspension with a reasonable concentration of
fibers. 

\begin{figure}[!tbp]
  \begin{center}
    \includegraphics[width=0.95\textwidth]{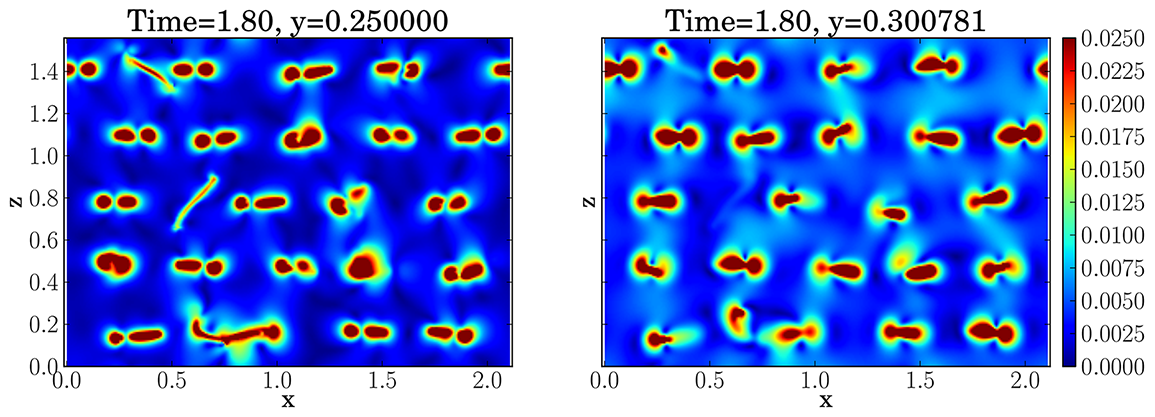} 
    \caption{Fluid deviation $\veldev$ on two horizontal planes for the
      $25$ fiber simulation computed in Table~\ref{Table:WeakScaling}.
      Plotted values are truncated at the threshold $\veldev = 0.025$.}
    \label{fig:FluidDeviationSlice} 
  \end{center}
\end{figure}

The next set of results attempts to quantify the importance of including the
full two-way fluid-structure interaction between fluid and fibers,
relative to other more common numerical approaches that simplify or
eliminate this interaction.  For this purpose, we define a quantity we
call the {\it local deviation} as 
\begin{gather*}
  \veldev(\bs{x},t) = \frac{| \bs{u}(\bs{x},t) - \bs{u}(\bs{x},0)
    |}{\max_{\bs{x}}(|\bs{u}(\bs{x},0)|)}, 
\end{gather*}
which is a local measure of the relative difference between the computed
fluid velocity and the corresponding linear shear flow that would arise
in the absence of any fibers.  We also define a related {\it global
  deviation} from linear shear using either the $\ell^\infty$-norm
\begin{gather*}
  \| \veldev(\bs{x},t) \|_\infty = \max_{i,j,k}
  |\veldev(\bs{x}_{i,j,k},t_n)|,
\end{gather*}
or $\ell^1$-norm
\begin{gather*}
  \| \veldev(\bs{x},t) \|_1 = \frac{h^3}{V} \sum_{i,j,k}
  |\veldev(\bs{x}_{i,j,k},t_n)|, 
\end{gather*}
where $V$ is the fluid volume.
For a 25-fiber simulation computed with $(P_x,P_y,P_z)=(5,1,5)$
processors, we provide plots in Figure~\ref{fig:FluidDeviationSlice} of
the local deviation $\veldev$ at time $t=1.80$ and along two different
horizontal slices.  The figures have truncated the values of $\veldev$
above the threshold 0.025 so that smaller deviations can be visualized.
From these plots we observe that the local deviation is largest adjacent
to the individual fibers where the no-slip condition forces the fluid to
follow the deforming and rotating fibers, but that the deviation decays
rapidly away from the fibers.  Nonetheless, there are still significant
fluid disturbances spread throughout the entire fluid domain that
influence fiber motion and are related to hydrodynamic interactions
between individual fibers.  The corresponding global deviation values
are $\|\veldev\|_1 = 0.0159$ and $\|\veldev\|_\infty = 0.135$ which show
that relative deviations in the flow are as high as 13.5\%\ near the
fibers but that the average over the entire flow field is only about
1.6\%.  Other simulations using different parameters and initial
conditions yield similar results (see \cite{Wiens2014}) with the average
relative deviation hovering around 2\%\ and the maximum ranging up to
40\%.  These results suggest that incorporating the full fluid-structure
interaction into models for non-dilute suspensions is important in
terms of properly capturing the dynamics of the flexible fibers.  We also
note that these simulations are performed at relative low values of
Reynolds number and fiber concentration, and that the deviation measure
will only get larger as the Reynolds number and concentration increase.

Finally, we close by investigating the parallel performance of our IB
algorithm by considering simulations of different-sized suspensions of
fibers on multiple processors. Based on our problem setup, the execution
time would ideally stay constant as the global problem size and number
of processors increase.  Indeed, Table~\ref{Table:WeakScaling} shows
that as the size of the fiber array ($P_x$, $P_z$) is increased, there
is only a slight increase in execution time and hence our algorithm is
said to be weakly scalable.  We remark that our code is still not fully
optimized and that the algorithm performance could be further improved
by making enhancements such as enforcing the top/bottom wall boundary
conditions directly instead of our approach of treating the walls using
IB tether points.

\begin{table}[htbp]\centering\small
  \caption{Weak scaling results showing the average execution time
    per time step (in seconds) for the multiple fiber problem.  The
    local problem size is held fixed as the number of processors $P$ (and
    global problem size) is increased.  Simulations are run on the
    Bugaboo cluster managed by WestGrid~\cite{Bugaboo}.} 
  \begin{tabular}{cc c}\toprule
    $P$  & ($P_x,P_y,P_z$)& Wall Time  \\
    \midrule
    $25$   &  ($5,1,5$)   & $0.57$ \\
    $64$   &  ($8,1,8$)   & $0.58$ \\
    $144$  &  ($12,1,12$) & $0.58$ \\
    $225$  &  ($15,1,15$) & $0.62$ \\
    $256$  &  ($16,1,16$) & $0.61$ \\
    \bottomrule
  \end{tabular}
  \label{Table:WeakScaling}
\end{table}

\section{Conclusions}
\label{sec:Conclusion}

In this paper, we have presented a parallel immersed boundary algorithm
for simulating suspensions of flexible fibers, where individual fibers
are modelled as Kirchhoff rods.  The novelty of this work derives from
its application to multi-fiber suspension flows with non-zero Reynolds
number and the inclusion of the the full two-way interaction between the
fluid and suspended fibers.  In our numerical simulations, we reproduce
the full range of orbital dynamics observed experimentally by Mason and
co-workers for isolated fibers immersed in a linear shear flow.  When
extending the results to multi-fiber suspensions, we demonstrate through
a weak scalability test that the parallel scaling of our algorithm is
near optimal and hence shows promise for simulating more complex
scenarios such as semi-dilute suspensions and fiber flocculation.

In the future, we plan to improve on the underlying model, which will
allow us to simulate more realistic fiber suspensions. First, we plan on
incorporating the contact forces between fibers such as the frictional
forces modelled by Schmid et al.~\cite{Schmid2000}.  Second, we will
incorporate the effect of added fiber mass using the penalty IB
method~\cite{Kim2007}.  After incorporating these extensions, a more
extensive comparison to experimental data would be required, comparing
quantities such as the specific viscosity of the
suspension~\cite{Petrie1999}.

\bibliography{Paper}

\begin{thebibliography}{10}

\bibitem{Arlov1958}
A.~P. Arlov, O.~L. Forgacs, and S.~G. Mason.
\newblock Particle motions in sheared suspensions\ \ {IV}.\ {G}eneral behaviour
  of wood pulp fibres.
\newblock {\em svenskpapp}, 61(3):61--67, 1958.

\bibitem{Batchelor1970}
G.~K. Batchelor.
\newblock Slender-body theory for particles of arbitrary cross-section in
  {S}tokes flow.
\newblock {\em Journal of Fluid Mechanics}, 44(3):419--440, 1970.

\bibitem{Bretherton1962}
F.~P. Bretherton.
\newblock The motion of rigid particles in a shear flow at low {R}eynolds
  number.
\newblock {\em Journal of Fluid Mechanics}, 14(2):284--304, 1962.

\bibitem{BringleyPeskin2008}
T.~T. Bringley and C.~S. Peskin.
\newblock Validation of a simple method for representing spheres and slender
  bodies in an immersed boundary method for {S}tokes flow on an unbounded
  domain.
\newblock {\em Journal of Computational Physics}, 227:5397--5425, 2008.

\bibitem{Cope1969}
O.~Cope.
\newblock Polymer blends of polyethylene terephthalate and
  alpha-olefin,alpha,beta-unsaturated carboxylic acid copolymers, March 25
  1969.
\newblock US Patent 3,435,093.

\bibitem{WolframAlphaDacron}
Dacron flexural modulus. {W}olfram{A}lpha.
\newblock Retrieved February 10, 2014 from
  \url{http://www.wolframalpha.com/input/?i=dacron+flexural+modulus}.

\bibitem{Dill1992}
E.~H. Dill.
\newblock Kirchhoff's theory of rods.
\newblock {\em Archive for History of Exact Sciences}, 44(1):1--23, 1992.

\bibitem{Tamdoo1981}
P.~A.~T. Doo and R.~J. Kerekes.
\newblock A method to measure wet fiber flexibility.
\newblock {\em Tappi Journal}, 64(3):113--116, 1981.

\bibitem{Tamdoo1982}
P.~A.~T. Doo and R.~J. Kerekes.
\newblock The flexibility of wet pulp fibres.
\newblock {\em Pulp and Paper Canada}, 83(2):46--50, 1982.

\bibitem{Forgacs1959}
O.~L. Forgacs and S.~G. Mason.
\newblock Particle motions in sheared suspensions {X}. orbits of flexible
  threadlike particles.
\newblock {\em Journal of Colloid Science}, 14:473--491, 1959.

\bibitem{Forgacs1958}
O.~L. Forgacs, A.~A. Robertson, and S.~G. Mason.
\newblock The hydrodynamic behaviour of paper-making fibres.
\newblock {\em Pulp Paper Magazine}, 59(5):117--128, 1958.

\bibitem{Griffith2012-2}
B.~E. Griffith and S.~Lim.
\newblock Simulating an elastic ring with bend and twist by an adaptive
  generalized immersed boundary method.
\newblock {\em Communications in Computational Physics}, 12(2):433, 2012.

\bibitem{Griffith2009}
B.~E. Griffith, X.~Y. Luo, D.~M. McQueen, and C.~S. Peskin.
\newblock Simulating the fluid dynamics of natural and prosthetic heart valves
  using the immersed boundary method.
\newblock {\em International Journal of Applied Mechanics}, 1(1):137--177,
  2009.

\bibitem{Griffith2005}
B.~E. Griffith and C.~S. Peskin.
\newblock On the order of accuracy of the immersed boundary method: {H}igher
  order convergence rates for sufficiently smooth problems.
\newblock {\em Journal of Computational Physics}, 208(1):75--105, 2005.

\bibitem{Guermond2010}
J.~L. Guermond and P.~D. Minev.
\newblock A new class of fractional step techniques for the incompressible
  {N}avier-{S}tokes equations using direction splitting.
\newblock {\em Comptes Rendus Mathematique}, 348:581--585, 2010.

\bibitem{Guermond2011}
J.~L. Guermond and P.~D. Minev.
\newblock A new class of massively parallel direction splitting for the
  incompressible {N}avier-{S}tokes equations.
\newblock {\em Computer Methods in Applied Mechanics and Engineering},
  200(23-24):2083--2093, 2011.

\bibitem{Hamlet2011}
C.~Hamlet, A.~Santhanakrishnan, and L.~A. Miller.
\newblock A numerical study of the effects of bell pulsation dynamics and oral
  arms on the exchange currents generated by the upside-down jellyfish
  \emph{{C}assiopea xamachana}.
\newblock {\em Journal of Experimental Biology}, 214:1911--1921, 2011.

\bibitem{Harlow1965}
F.~H. Harlow and J.~E. Welch.
\newblock Numerical calculation of time-dependent viscous incompressible flow
  of fluid with free surface.
\newblock {\em Physics of Fluids}, 8(12):2182--2189, 1965.

\bibitem{Jeffery1922}
G.~B. Jeffery.
\newblock The motion of ellipsoidal particles immersed in a viscous fluid.
\newblock {\em Proceedings of the Royal Society of London. Series A},
  102(715):161--179, 1922.

\bibitem{Joung2001}
C.~G. Joung, N.~Phan-Thien, and X.~J. Fan.
\newblock Direct simulation of flexible fibers.
\newblock {\em Journal of Non-Newtonian Fluid Mechanics}, 99(1):1--36, 2001.

\bibitem{keshtkar-etal-2009}
M.~Keshtkar, M.~C. Heuzey, and P.~J. Carreau.
\newblock Rheological behavior of fiber-filled model suspensions: {E}ffect of
  fiber flexibility.
\newblock {\em Journal of Rheology}, 53(3):631--650, 2009.

\bibitem{Kim2007}
Y.~Kim and C.~S. Peskin.
\newblock Penalty immersed boundary method for an elastic boundary with mass.
\newblock {\em Physics of Fluids}, 19(5):053103, 2007.

\bibitem{Kim2009}
Y.~Kim and C.~S. Peskin.
\newblock 3{D} parachute simulation by the immersed boundary method.
\newblock {\em Computers \& Fluids}, 38(6):1080--1090, 2009.

\bibitem{Kim2012}
Y.~Kim, Y.~Seol, M.~C. Lai, and C.~S. Peskin.
\newblock The immersed boundary method for two-dimensional foam with
  topological changes.
\newblock {\em Communications in Computational Physics}, 12(2):479, 2012.

\bibitem{Lai2000}
M.~C. Lai and C.~S. Peskin.
\newblock An immersed boundary method with formal second-order accuracy and
  reduced numerical viscosity.
\newblock {\em Journal of Computational Physics}, 160(2):705--719, 2000.

\bibitem{Li2013}
L.~Li, H.~Manikantan, D.~Saintillan, and S.~E. Spagnolie.
\newblock The sedimentation of flexible filaments.
\newblock {\em Journal of Fluid Mechanics}, 735:705--736, 2013.

\bibitem{Lim2010}
S.~Lim.
\newblock Dynamics of an open elastic rod with intrinsic curvature and twist in
  a viscous fluid.
\newblock {\em Physics of Fluids}, 22(2):024104--024104, 2010.

\bibitem{Lim2008}
S.~Lim, A.~Ferent, X.~S. Wang, and C.~S. Peskin.
\newblock Dynamics of a closed rod with twist and bend in fluid.
\newblock {\em SIAM Journal on Scientific Computing}, 31(1):273--302, 2008.

\bibitem{Lindstrom2007}
S.~B. Lindstr{\"o}m and T.~Uesaka.
\newblock Simulation of the motion of flexible fibers in viscous fluid flow.
\newblock {\em Physics of Fluids}, 19(11):113307, 2007.

\bibitem{Lindstrom2008}
S.~B. Lindstr{\"o}m and T.~Uesaka.
\newblock Simulation of semidilute suspensions of non-brownian fibers in shear
  flow.
\newblock {\em Journal of Chemical Physics}, 128(2):024901, 2008.

\bibitem{Lindstrom2009}
S.~B. Lindstr{\"o}m and T.~Uesaka.
\newblock A numerical investigation of the rheology of sheared fiber
  suspensions.
\newblock {\em Physics of Fluids}, 21:083301, 2009.

\bibitem{Mori2008}
Y.~Mori and C.~S. Peskin.
\newblock Implicit second-order immersed boundary methods with boundary mass.
\newblock {\em Computer Methods in Applied Mechanics and Engineering},
  197(25–28):2049--2067, 2008.

\bibitem{Nguyen2014}
H.~Nguyen and L.~Fauci.
\newblock Hydrodynamics of diatom chains and semiflexible fibres.
\newblock {\em Journal of The Royal Society Interface}, 11(96):20140314, 2014.

\bibitem{Olson2013}
S.~D. Olson, S.~Lim, and R.~Cortez.
\newblock Modeling the dynamics of an elastic rod with intrinsic curvature and
  twist using a regularized {S}tokes formulation.
\newblock {\em Journal of Computational Physics}, 2013.

\bibitem{Peskin1972}
C.~S. Peskin.
\newblock {Flow Patterns Around Heart Valves: A Numerical Method}.
\newblock {\em Journal of Computational Physics}, 10:252--271, 1972.

\bibitem{Peskin2002}
C.~S. Peskin.
\newblock The immersed boundary method.
\newblock {\em Acta Numerica}, 11:479--517, 2002.

\bibitem{Petrie1999}
C.~J.~S. Petrie.
\newblock The rheology of fibre suspensions.
\newblock {\em Journal of Non-Newtonian Fluid Mechanics}, 87(2):369--402, 1999.

\bibitem{Rejniak2007}
K.~A. Rejniak and R.~H. Dillon.
\newblock A single cell-based model of the ductal tumour microarchitecture.
\newblock {\em Computational and Mathematical Methods in Medicine},
  8(1):51--69, 2007.

\bibitem{Ross1997}
R.~F. Ross and D.~J. Klingenberg.
\newblock Dynamic simulation of flexible fibers composed of linked rigid
  bodies.
\newblock {\em Journal of Chemical Physics}, 106:2949--2960, 1997.

\bibitem{Schmid2000}
C.~F. Schmid, L.~H. Switzer, and D.~J. Klingenberg.
\newblock Simulations of fiber flocculation: Effects of fiber properties and
  interfiber friction.
\newblock {\em Journal of Rheology}, 44(3):781--809, 2000.

\bibitem{Skjetne1997}
P.~Skjetne, R.~F. Ross, and D.~J. Klingenberg.
\newblock Simulation of single fiber dynamics.
\newblock {\em Journal of {C}hemical {P}hysics}, 107:2108--2121, 1997.

\bibitem{Stockie1997}
J.~M. Stockie.
\newblock {\em Analysis and Computation of Immersed Boundaries, with
  Application to Pulp Fibres}.
\newblock PhD thesis, Institute of Applied Mathematics, University of British
  Columbia, Vancouver, Canada, 1997.
\newblock Available from \url{https://circle.ubc.ca/handle/2429/7346}.

\bibitem{Stockie2002}
J.~M. Stockie.
\newblock Simulating the dynamics of flexible wood pulp fibres in suspension.
\newblock In {\em {P}roceedings of the 16th {A}nnual {I}nternational
  {S}ymposium on {H}igh {P}erformance {C}omputing {S}ystems and
  {A}pplications}, page 154. IEEE Computer Society, 2002.

\bibitem{Stockie1998}
J.~M. Stockie and S.~I. Green.
\newblock Simulating the motion of flexible pulp fibres using the immersed
  boundary method.
\newblock {\em Journal of Computational Physics}, 147(1):147--165, 1998.

\bibitem{Switzer2004}
L.~H. Switzer and D.~J. Klingenberg.
\newblock Flocculation in simulations of sheared fiber suspensions.
\newblock {\em International Journal of Multiphase Flow}, 30(1):67--87, 2004.

\bibitem{switzer-klingenberg-2003}
L.~H. {Switzer III} and D.~J. Klingenberg.
\newblock Rheology of sheared flexible fiber suspensions via fiber-level
  simulations.
\newblock {\em Journal of Rheology}, 47(3):759--778, 2003.

\bibitem{TornbergShelley2004}
A.~K. Tornberg and M.~J. Shelley.
\newblock Simulating the dynamics and interactions of flexible fibers in
  {S}tokes flows.
\newblock {\em Journal of Computational Physics}, 196:8--40, 2004.

\bibitem{Wang2006}
G.~Wang, W.~Yu, and C.~Zhou.
\newblock Optimization of the rod chain model to simulate the motions of a long
  flexible fiber in simple shear flows.
\newblock {\em European Journal of Mechanics-B/Fluids}, 25(3):337--347, 2006.

\bibitem{Bugaboo}
WestGrid.
\newblock Quick{S}tart {G}uide to {B}ugaboo.
\newblock Retrieved April 11, 2013 from
  \url{http://www.westgrid.ca/support/quickstart/bugaboo}.

\bibitem{Wherrett1997}
G.~Wherrett, I.~Gartshore, M.~Salcudean, and J.~Olson.
\newblock A numerical model of fibre motion in shear.
\newblock In {\em Proceedings of the 1997 {ASME} Fluids Engineering Division
  Summer Meeting}, Vancouver, Canada, June 22--26, 1997.

\bibitem{White2006}
F.~M. White.
\newblock {\em Viscous fluid flow}, volume~46.
\newblock McGraw-Hill Higher Education Boston, 2006.

\bibitem{Wiens2014}
J.~K. Wiens.
\newblock {\em An efficient parallel immersed boundary algorithm, with
  application to the suspension of flexible fibers}.
\newblock PhD thesis, Department of Mathematics, Simon Fraser University,
  Burnaby, Canada, 2014.

\bibitem{Wiens2013}
J.~K. Wiens and J.~M. Stockie.
\newblock An efficient parallel immersed boundary algorithm using a
  pseudo-compressible fluid solver.
\newblock Submitted to Journal of Computational Physics. Preprint available at
  \url{http://arxiv.org/abs/1305.3976}, May 2013.

\bibitem{Wu2010-2}
J.~Wu and C.~K. Aidun.
\newblock A method for direct simulation of flexible fiber suspensions using
  lattice boltzmann equation with external boundary force.
\newblock {\em International Journal of Multiphase Flow}, 36(3):202--209, 2010.

\bibitem{Wu2010}
J.~Wu and C.~K. Aidun.
\newblock Simulating {3D} deformable particle suspensions using lattice
  {B}oltzmann method with discrete external boundary force.
\newblock {\em International {J}ournal for {N}umerical {M}ethods in {F}luids},
  62(7):765--783, 2010.

\bibitem{Yamamoto1993}
S.~Yamamoto and T.~Matsuoka.
\newblock A method for dynamic simulation of rigid and flexible fibers in a
  flow field.
\newblock {\em Journal of Chemical Physics}, 98:644, 1993.

\end{thebibliography}
\bibliographystyle{abbrv}

\end{document}